\documentclass[12pt]{article}

\usepackage{epsfig}
\usepackage{graphicx}
\usepackage{amssymb,amsmath}
\usepackage{hyperref}
\usepackage[nosort]{cite}
\hypersetup{
    colorlinks=true,
    linkcolor=blue,
    citecolor=red,
}

\usepackage{array}

\DeclareMathOperator\arctanh{arctanh}

\begin{document}

\begin{center}
\Large{\Large \bf Effects of quantum corrections to Lorentzian vacuum transitions in the presence of gravity} \vspace{0.7cm}

\large  H. Garc\'{\i}a-Compe\'an$^{*,a,}$\footnote{e-mail address: {\tt
		hugo.compean@cinvestav.mx}, $^*$Corresponding Author}, J. Hern\'andez-Aguilar$^{b,}$\footnote{e-mail address: {\tt
jorge.hernandezag@alumno.buap.mx}}, D. Mata-Pacheco$^{a,}$\footnote{e-mail
address: {\tt daniel.mata@cinvestav.mx}}, \ \ C. Ram\'{\i}rez$^{b,}$\footnote{e-mail
address: {\tt cramirez@fcfm.buap.mx}}

\vspace{0.3cm}
{\small \em $^a$Departamento de F\'{\i}sica, Centro de
	Investigaci\'on y de Estudios Avanzados del IPN}\\
{\small\em P.O. Box 14-740, CP. 07000, Ciudad de M\'exico, M\'exico}\\

\vskip .4truecm

{\small \em $^b$Facultad de Ciencias F\'{\i}sico Matem\'aticas, \\ 
Benem\'erita Universidad Aut\'noma de Puebla\\
4 Sur 104, Puebla 72000, Puebla, M\'exico}\\
\vspace{0.3cm}

\vspace*{1.5cm}
\end{center}

\begin{abstract}

 \vskip 1truecm
 We present a study of the vacuum transition probabilities taking into account quantum corrections. We first introduce a general method that expands previous works employing the Lorentzian formalism of the Wheeler-De Witt equation by considering higher order terms in the semiclassical expansion. The method presented is applicable in principle to any model in the minisuperspace and up to any desired order in the quantum correction terms. Then, we apply this method to obtain analytical solutions for the probabilities up to second quantum corrections for homogeneous isotropic and anisotropic universes. We use the Friedmann-Lemaitre-Robertson-Walker metric with positive and zero curvature for the isotropic case and the Bianchi III and Kantowski-Sachs metrics for the anisotropic case. Interpreting the results as distribution probabilities of creating universes by vacuum decay with a given size, we found that the general behaviour is that considering up to the second quantum correction leads to an avoidance of the initial singularity. However, we show that this result can only be achieved for the isotropic universe. Furthermore, we also study the effect of anisotropy on the transition probabilities.

\end{abstract}

\bigskip

\newpage

\section{Introduction}
\label{S-Intro}
Quantum tunnelling represents one of the first phenomena that challenged the classical ideas and required a quantum theory for its proper understanding. It is a key process in many situations, in particular, in the canonical approach to quantum gravity it is thought to provide the description for the birth of our universe. Let us consider a general scenario where a scalar field is present with its corresponding potential with the general behaviour of containing two local minima of different values separated by a hill. Then, we can always expect that a transition between the two minima can occur via quantum tunnelling. In field theory, this is called a vacuum transition (analogously in the absence of a scalar field,  the two minima can also be described in a simpler context with two different values of the cosmological constant). Since this set up appears in many scenarios, it has attracted a lot of attention over the decades. In particular, for field theory, the study of such types of transitions began with the work of Sidney Coleman et. al. \cite{Coleman:1977py,Callan:1977pt} and it is described by the nucleation of true vacuum bubbles on a false vacuum background (the true vacuum in this context is located at the global minimum). Later on, by using a proposal for a quantum theory of gravity employing the path integral approach, these transitions were studied including the gravitational field by Coleman and De Luccia \cite{Coleman:1980aw} and then by Parke \cite{Parke:1982pm}. Over the years, many new results have been explored using these Euclidean techniques (see for example \cite{Jensen:1989fh,Samuel:1991dy,Wu:1992ht,Mansouri:1992er,Gen:1999gi,Cai:2008ht,Kanno:2011vm,Salehian:2018yoq,Oshita:2021aux,Vicentini:2022opn,Antoniadis:2024ent}).

On the other hand, it is well known that an alternative description to the path integral approach of quantum gravity can be described by employing a Hamiltonian formalism. In this way, employing the Arnowitt, Deser and Misner (ADM) formalism \cite{Arnowitt:1962hi,Corichi:1991qqo}, instead of a path integral, the classical Hamiltonian and momentum constraints of a gravitational system are quantized using the Dirac quantization procedure, leading to a set of constraints on a wave function for the universe (for a general introduction to the subject of quantum cosmology see \cite{Halliwell:2009,Wiltshire:1995vk,Carlip:2001wq,Kiefer:2008sw,Isham:1992ms}). In particular, when we are dealing with homogeneous metrics, all the information is encoded in the equation resulting from the Hamiltonian constraint. At the quantum level, this constraint can be regarded as the Hamiltonian differential operator acting on a wave function, this is called the Wheeler-De Witt (WDW) equation \cite{Wheeler,DeWitt} and the wave function is normally termed the wave function of the universe. One important feature of this formalism is that it does not rely on a Wick rotation, thus, it is a purely Lorentzian method \footnote{In this particular context what we mean by this statement is that we will be able to study the transition probabilities without requiring any Wick rotation whereas in the Euclidean path integral approach an analytic continuation is required in order to describe the universe after nucleation which may be a relevant limitation.}. The interpretation of this wave function turned out to be very troublesome, since in a cosmological setting describing the entire universe, there is no notion of an external observer (see for example \cite{Craig:2010vf,Chataignier:2023rkq}  for recent proposals and discussions on this regard). However, there are many ways in which we can use the solutions of the WDW equation to explore physical properties of the gravitational systems. In particular, the squared ratio of two solutions can be interpreted as the conditional probability between both configurations. Using this interpretation, the vacuum transition probabilities have also been explored in a Lorentzian formalism. First of all, Fischler, Morgan and Polchinski studied the transition between two values of the cosmological constant in \cite{FMP1,FMP2}. Recently, such results were generalized in \cite{deAlwis:2019dkc}, one important feature of the Lorentzian formalism is that the final state can be a closed universe, contrary to the results obtained with the Euclidean approach, that is employing the Euclidean path integral. Later on, the inclusion of a scalar field was studied in \cite{Cespedes:2020xpn}. One limitation of this formalism when dealing with the scalar field is that there is not a concrete description of bubble nucleation, however the transition probabilities obtained can be interpreted as probability distributions of creating universes with a given size by vacuum decay, as a generalization of the standard tunnelling from nothing scenario. Based on these works, in \cite{Garcia-Compean:2021syl} a general method was proposed to compute such transition probabilities for any model in the minisuperspace at the semiclassical level. In \cite{Garcia-Compean:2021vcy} such method was used to study the transitions in the Ho\v{r}ava-Lifshitz theory of gravity, then in \cite{Garcia-Compean:2022ysy} it was generalized to consider linear terms in the momenta to take into account the effect of a Generalized Uncertainty Principle. 

Let us remark that all these studies provide results only at the semiclassical level. The incorporation of quantum corrections to the transition probabilities in the Euclidean formalism is troublesome since there appears a negative mode problem that may spoil the semiclassical approximation \cite{Lavrelashvili:1985vn}. This issue has been studied rigorously over the years \cite{Abbott:1987xq,Lavrelashvili:1999sr,Koehn:2015hga,Bramberger:2019mkv,Jinno:2020zzs}. On the other hand, the Lorentzian formalism employs a semiclassical expansion in the form of a WKB proposal. In this way, the quantum corrections are incorporated by considering the higher-order terms in the $\hbar$ expansion. It has been shown for field theory that the first quantum correction computed in this form indeed coincides with the 1-loop contribution of the Euclidean formalism \cite{deAlwis:2023gth}. Thus, the main purpose of this article is to expand the general method presented in \cite{Cespedes:2020xpn,Garcia-Compean:2021syl} to incorporate quantum corrections to the transition probabilities with gravity in a general setup, in this way avoiding the problems presented in the Euclidean formalism. For recent developments regarding the incorporation of quantum gravity corrections employing the WDW equation see for example  \cite{DiGioia:2019nti,Maniccia:2021skz,Maniccia:2023cgv,Maniccia:2023fkt,Maniccia:2024bax}.

Let us also remark that recently, a new method has been proposed (at the semiclassical level) in the form of a tunnelling potential in which many of the Euclidean results can be derived  \cite{Espinosa:2018voj,Espinosa:2021tgx,Calcagni:2022tls,Espinosa:2022ofv,Espinosa:2022jlx}. Furthermore, the holographic interpretation of these transitions has also been explored \cite{Freivogel:2006xu,Barbon:2010gn,Ghosh:2021lua,Pasquarella:2022ibb}. In this way, the computation of quantum corrections obtained in the present work may be relevant in the holographic setup. Moreover, if the new Euclidean method is able to compute loop corrections as well, a possible comparison with our results is highly relevant as well.

Moreover, although the  Friedmann-Lemaitre-Robertson-Walker (FLRW) metric describing a homogeneous and isotropic universe is the natural and most used choice to study these transitions given its cosmological importance, it is also relevant to study homogeneous but anisotropic metrics. This is justified by the fact that inflation is thought to erase any signal of anisotropy in the early universe. However, in the very early universe certain amount of anisotropy could be present and such metrics should be relevant. In addition, there are also some recent experimental studies that suggest that such metrics may be important for the description of our universe even at the present epoch as well \cite{Colin:2018ghy,Migkas:2020fza}. Therefore, in this article, we will apply the general method to study both isotropic and anisotropic homogeneous metrics.

The outline of this article is as follows: In Section \ref{S-KSNormal} we will present the general method to compute the transition probabilities with quantum corrections for any model on the minisuperspace starting with a generic form of the Hamiltonian constraint. We will apply the method to study the transition probabilities with quantum corrections terms (up to second order in powers of $\hbar$) in the following sections. First, we will study the isotropic case with the closed FLRW metric in Section \ref{S-PFLRW} and the flat FLRW metric in Section \ref{S-FFLRW}. Then, we will consider anisotropic metrics. In Section \ref{S-B3} we will study Bianchi III metric and compare the results with the flat FLRW result, which corresponds to its isotropy limit. Then, in Section \ref{S-KS} we will use the Kantowski-Sachs metric. Finally, in Section \ref{S-FinalRemarks} we will present our Final Remarks and in Appendix \ref{S-AP}, we will discuss the dependence of the results on the choice of factorization for the Bianchi III and flat FLRW metrics which is an important subject to remark.
\section{Transition probabilities with quantum corrections}
\label{S-KSNormal} 
In this section, we begin by expanding the general method proposed in \cite{Cespedes:2020xpn, Garcia-Compean:2021syl} to compute transition probabilities in a general scenario. We will consider higher-order terms in the WKB expansion for the WDW equation to compute the quantum corrections to the transition probabilities. We will also consider transitions between a false and a true vacuum of a scalar field potential. We employ the notation and conventions of these references. 

We consider a general form for the Hamiltonian constraint in the context of the ADM formulation of general relativity given by 
        \begin{equation}\label{HamConst}
             \mathcal{H}=\frac{1}{2}G^{M N}(\Phi)\pi_{M}\pi_{N}+f[\Phi] \simeq 0,
        \end{equation}
where Wheeler's superspace is defined by the coordinates $\Phi^{M}$ (there can be an infinite number of them and include the degrees of freedom of the three-dimensional metric as well as matter fields and are collectively denoted by $\Phi$ with their corresponding canonically conjugate momenta $\pi_M$). Moreover, $G^{M N}$ represents the superspace inverse metric, and the function $f[\Phi]$ will contain all remaining terms related to the three-curvature and potential terms of matter fields. Hence, in general, these are 3-fields $\Phi^{M}(\vec{x})$, and the corresponding degrees of freedom are labelled by the index $M$ and the 3-vector $\vec{x}$. The WDW equation is obtained by performing a canonical quantization of the Hamiltonian constraint by promoting the superspace degrees of freedom and their momenta to hermitian operators acting on the wave functional. Thus, in the position representation, the wave functional is an eigenfunction of the superspace operators $\widehat{\Phi}^{M}(\vec{x})$, which have eigenvalues $\Phi^{M}(\vec{x})$ in specific real domains, for given $M$ and $\vec{x}$, and the corresponding momenta operators are $\widehat{\pi}_{M}= -i \hbar \frac{\delta}{\delta \Phi^{M}(\vec{x})}$, leading to the Hamiltonian constraint
        \begin{equation}
             \mathcal{H} \Psi(\Phi)=\left[-\frac{\hbar^{2}}{2} G^{M N}(\Phi) \frac{\delta}{\delta \Phi^{M}} \frac{\delta}{\delta \Phi^{N}}+f[\Phi]\right] \Psi[\Phi]=0, 
         \label{eq:WDWGR}
        \end{equation}
up to ordering ambiguities\footnote{A natural ordering choice would be to take the hermitian ordering ansatz $G^{M N}(\Phi)\pi_{M}\pi_{N}\to\widehat{\pi}_{M}G^{M N}(\widehat{\Phi})\widehat{\pi}_{N}$. However, the ordering that we are taking is useful for the purposes of the current work.}, where $\Psi[\Phi]$ is the wave functional in superspace. In order to obtain a semiclassical outcome and quantum corrections from this equation, we employ the general proposal of the WKB type
        \begin{equation}
            \Psi[\Phi]=\exp \left\{\frac{i}{\hbar} S[\Phi]\right\}
        \end{equation}
with the $\hbar$-expansion
        \begin{equation}
             S[\Phi]=S_{0}[\Phi]+\hbar S_{1}[\Phi]+\hbar^2 S_{2}[\Phi]+\mathcal{O}\left(\hbar^{3}\right) ,
          \label{eq:WKBExpan}
        \end{equation}
where $S_{0}$ is the classical action and $S_{1}$ and $S_{2}$ are the first and second quantum corrections respectively. Substituting this ansatz in the WDW equation (\ref{eq:WDWGR}) we obtain for the first three orders in $\hbar$
        \begin{equation}
             \frac{1}{2} G^{M N} \frac{\delta S_{0}}{\delta \Phi^{M}} \frac{\delta S_{0}}{\delta \Phi^{N}}+f[\Phi]=0,
            \label{eq:WKB3O}
        \end{equation}
        \begin{equation}
             2 G^{M N} \frac{\delta S_{0}}{\delta \Phi^{M}} \frac{\delta S_{1}}{\delta \Phi^{N}}=i G^{M N} \frac{\delta^{2}}{\delta \Phi^{M} \delta \Phi^{N}} S_{0},
        \label{eq:WKB1}
        \end{equation}
        \begin{equation}
            2 G^{M N} \frac{\delta S_0}{\delta \Phi^M} \frac{\delta S_2}{\delta \Phi^N}+G^{M N} \frac{\delta S_1}{\delta \Phi^M} \frac{\delta S_1}{\delta \Phi^N}=i G^{M N} \frac{\delta^2 S_1}{\delta \Phi^M \delta \Phi^N}.
         \label{eq:WKB2}
        \end{equation}  
We now assume that the superspace has a finite dimension $n$, that is we make a minisuperspace approximation where there are $n$ coordinates on the minisuperspace $\Phi^{M}$. The solution of equation (\ref{eq:WKB3O}) is the classical action $S_0$. This action generates classical trajectories in minisuperspace, i.e. integral curves $\Phi_s$ with parameter $s$, along the gradient of $S_0$, which can be given as solutions of the equation  
        \begin{equation}
            C(s) \frac{d \Phi^{M}_{s}}{d s}=G^{M N} \frac{\delta S_{0}}{\delta \Phi^{N}_{s}} ,
            \label{eq:Cs}
        \end{equation}
where $C(s)$ is an auxiliary function, an explanation for the introduction of this function was given in \cite{Garcia-Compean:2021syl}. Then, employing eqs. (\ref{eq:WKB3O}) and (\ref{eq:Cs}) the classical action can be written as
        \begin{equation}
            S_{0}\left[\Phi\right]=-2 \int^{s} \frac{d s^{\prime}}{C\left(s^{\prime}\right)} \int_{X} f\left[\Phi_{s^{\prime}}\right].
            \label{eq:S0f}
        \end{equation}
Then, as it was shown in \cite{Garcia-Compean:2021syl} a system of equations for the $n+1$ variables: $\left(\frac{d \Phi^{M}_{s}}{d s}, C^{2}(s)\right)$ can be obtained which in principle can be solved for any model in the minisuperspace in order to obtain a solution for $S_{0}$. In the following, we will assume that the fields $\Phi^{M}_{s}$ depend only on the time variable, then using eq. (\ref{eq:S0f}), the functional derivative in (\ref{eq:Cs}) can be expressed in terms of a partial derivative of the function $f[\Phi]$ and the integrals on the spatial slice will only give a factor of the volume of such slice $\operatorname{Vol}(X)$.

In this way, considering the next order in the WKB expansion, we note that we can write
         \begin{equation}
            \frac{d S_{1}}{d s}=\int_{X} \frac{d \Phi^{M}_{s}}{d s} \frac{\delta S_{1}}{\partial \Phi^{M}_{s}}, 
        \end{equation}
then, using (\ref{eq:Cs}), from eq. (\ref{eq:WKB1}) we obtain
        \begin{equation}
              S_{1}=\frac{i}{2} \int^{s} \int_{X} \frac{d s'}{C(s')} G^{M N} \frac{\delta^{2} S_{0}}{\delta \Phi^{M}_{s'} \delta \Phi^{N}_{s'}}.
        \end{equation}
Furthermore, from (\ref{eq:S0f}), it is clear that we can write
        \begin{equation}
             \frac{\delta S_{0}}{\delta \Phi^{M}_{s}}=-\frac{2 \operatorname{Vol}(X)}{C(s)} \frac{\partial f}{\partial \Phi^{M}_{s}},
        \end{equation}
and 
        \begin{equation}
            \frac{\delta^{2} S_{0}}{\delta \Phi^{M}_{s} \delta \Phi^{N}_{s}}=-\frac{2 \operatorname{Vol}(X)}{C(s)} \frac{\partial^2 f}{\partial \Phi^{M}_{s} \partial \Phi^{N}_{s}}.
        \end{equation}
Thus, it is found that the first quantum correction can be written as
        \begin{equation}
            S_{1}=-i \operatorname{Vol}^{2}(X) \int^{s} \frac{ds'}{C^{2}(s')} \nabla^2 f ,
            \label{eq:S1}
        \end{equation}
where the Laplacian is defined on the minisuperspace, that is
	\begin{equation}
		\nabla^{2} f=G^{M N} \frac{\partial^2 f}{\partial \Phi^{M} \partial \Phi^{N}} .
	\end{equation}

Likewise, following the same procedure using (\ref{eq:WKB2}), we obtain that the second quantum correction can be written as
        \begin{equation}
        	S_2=\frac{1}{2} \int^{s} ds' \frac{\operatorname{Vol}^3(X)}{C^3(s')}\nabla^{2}\left(\nabla^{2} f\right)           +\frac{1}{2} \int^{s} ds' \frac{\operatorname{Vol}^5(X)}{C^5(s')} \left[\nabla\left(\nabla^{2} f\right)\right]^{2},
        	\label{eq:S2}
        \end{equation}       
where we have denoted
        \begin{equation}
        	\quad(\nabla f)^{2}=G^{M N} \frac{\partial f}{\partial \Phi^{M}} \frac{\partial f}{\partial \Phi^{N}}.
        \end{equation}
Thus, we note that once we have solved the system of equations, using (\ref{eq:S0f}), (\ref{eq:S1}) and (\ref{eq:S2}) we can have a solution to the WDW equation up to second order in the WKB expansion in general. Let us also remark that each term in the WKB expansion leads to an independent equation from the WDW equation (such as eqs. (\ref{eq:WKB3O})-(\ref{eq:WKB2}) for the first three terms), thus, the system of equations will always have the same number of variables as equations. Then, in principle, we can compute the transition probabilities up to any desired order. However, for simplicity, in the following we will keep only up to the second quantum correction. 
  
As we remarked in \cite{Garcia-Compean:2021syl}, the system of equations can be solved in general as long as $\Phi^{M}=\Phi^{M}(s)$ is satisfied, yielding the solutions
	\begin{equation}
		C^{2}(s)=-\frac{2 \operatorname{Vol}^{2}(X)}{f}(\nabla f)^{2}, \quad \frac{d \Phi^{M}_{s}}{d s}=\frac{f}{\operatorname{Vol}(X)} \frac{\nabla^{M} f}{(\nabla f)^{2}}, 
	\label{eq:Cs2}
	\end{equation}
where we rise indices with the minisuperspace metric, that is
\begin{equation}
\nabla^{M} f=G^{M N} \frac{\partial f}{\partial \Phi^{N}} .
\end{equation}

From eq. (\ref{eq:Cs2}) we see that in general the various fields will be related by
\begin{equation}\label{Relations}
	\frac{d \Phi^{M}}{d \Phi^{N}}=\frac{\nabla^{M}f}{\nabla^{N}f},
\end{equation}
which is valid for every value of $M$ and $N$ such that $d \Phi^{M, N} \neq 0$. Therefore, as long as all the expressions are different from zero, we will always be able to reduce the number of independent degrees of freedom to just one and change the integration variable from $s$ to such field in (\ref{eq:S0f}), (\ref{eq:S1}) and (\ref{eq:S2}).

So far, we have only studied the way in which we can obtain a general solution from the WDW equation with up to second quantum correction terms. In order to compute the transition probabilities, following \cite{Cespedes:2020xpn,Garcia-Compean:2021syl} we will consider the ratio of two solutions, in one of them we follow a path in field space in which the scalar field evolves from the false minimum to the true one, whereas in the other one, we discuss a solution where the scalar field is kept at the false minimum. Then, the transition tunnelling probability for going from the false vacuum at $\phi_{A}$ to the true vacuum at $\phi_{B}$ is given by
        \begin{equation}
            P(A \rightarrow B)=\exp [-2 \operatorname{Re}(\Gamma)],
         \label{eq:Prob}
        \end{equation}
where 
	\begin{equation}
		\pm \Gamma=\frac{i}{\hbar}\big[S\left(\varphi_{0}^{I}, \phi_{B} ; \varphi_{m}^{I}, \phi_{A}\right)-S\left(\varphi_{0}^{I}, \phi_{A} ; \varphi_{m}^{I}, \phi_{A}\right)\big] ,
	\end{equation}
and the degrees of freedom coming from the three metric are denoted by $\varphi^{I}$, these fields take an initial value of  $\varphi^{I}(s=0)=\varphi_{0}^{I}$ where the scalar field is $\phi_{B}$ and a final value of $\varphi^{I}\left(s=s_{M}\right)=\varphi_{m}^{I}$ where the scalar field is $\phi_{A}$. Moreover,  the sign ambiguity appears due to the fact that the general solution of the wave functionals will be a linear superposition of exponential terms, however, we will keep only the dominant terms. Thus, using the expansion  (\ref{eq:WKBExpan}) we can write in this case up to second order
        \begin{equation}\label{DefGammaGe}
             \pm \Gamma=\Gamma_{0}+\Gamma_{1}+\Gamma_{2},
        \end{equation}
where 
        \begin{equation}
            \begin{array}{l}
            \Gamma_0=\frac{i}{\hbar}{\big[S_0\left(\varphi_0^I, \phi_B ; \varphi_m^I, \phi_A\right)-S_0\left(\varphi_0^I, \phi_A ; \varphi_m^I, \phi_A\right)\big]}, \\
            \Gamma_1= i\big[S_1\left(\varphi_0^I, \phi_B ; \varphi_m^I, \phi_A\right)-S_1\left(\varphi_0^I, \phi_A ; \varphi_m^I, \phi_A\right)\big],\\
            \Gamma_2=i \hbar\big[S_2\left(\varphi_0^I, \phi_B ; \varphi_m^I, \phi_A\right)-S_2\left(\varphi_0^I, \phi_A ; \varphi_m^I, \phi_A\right)\big].
            \end{array}
            \end{equation}
In this case, $\Gamma_{0}$ stands for the value that is calculated with the classical action, and it will be referred to as the semiclassical contribution, $\Gamma_{1} $ is the first quantum correction to the transition probabilities, and $\Gamma_{2}$ is the second quantum correction.

As it was remarked in  \cite{Cespedes:2020xpn,Garcia-Compean:2021syl}, the concrete expression for the transition probabilities will depend on the specific choice of $s$. In this work we will consider the same form as in previous works, that is we consider the parameter $s$ such that the scalar field takes the values
\begin{equation}
	\phi(s) \approx \begin{cases}\phi_B, & 0<s<\bar{s}-\delta s \\ \phi_A, & \bar{s}+\delta s<s<s_M.\end{cases}
	\label{eq:sparameter}
\end{equation}
Thus, we establish a configuration in which there is a bubble of true vacuum in the background of a false vacuum separated by a wall. Given the limitations of the minisuperspace approach in describing these transitions, this method remains our sole option for implementing this configuration. Furthermore, we will always consider the thin wall approximation in which $\delta s \to0$. This choice for the $s$ parameter allows us to obtain the solutions from the semiclassical computations on the Euclidean approach as was shown in \cite{Cespedes:2020xpn,Garcia-Compean:2021syl}.

With this choice, from (\ref{eq:S0f}) we obtain that the semiclassical contribution takes the form
        \begin{equation}
         \begin{aligned}
            \Gamma_0= & -\frac{2 \operatorname{Vol}(X) i}{\hbar}\left\{\left.\int_{0}^{\bar{s}-\delta s} \frac{d s}{C(s)} f\right|_{\phi=\phi_B}-\left.\int_{0}^{s-\delta s} \frac{d s}{C(s)} f\right|_{\phi=\phi_A}\right. \\
             & \left.+\int_{\bar{s}-\delta s}^{\bar{s}+\delta s} d s\left[\frac{f}{C(s)}-\left.\frac{f}{C(s)}\right|_{\phi=\phi_A}\right]\right\}.
         \label{eq:gammazero}
        \end{aligned}
        \end{equation}
From (\ref{eq:S1}) the first quantum correction is given by
        \begin{equation}
          \begin{aligned}
            \Gamma_{1}& = \operatorname{Vol}^{2}(X)\left[\int_{0}^{\bar{s}-\delta s}\right. \left.\frac{d s}{C^{2}(s)} \nabla^2f \right|_{\phi=\phi_{B}}-\left.\int_{0}^{\bar{s}-\delta s} \frac{d s}{C^{2}(s)} \nabla^2f\right|_{\phi=\phi_{A}} \\
             &\left. +\int_{\bar{s}-\delta s}^{\bar{s}+\delta s} ds\frac{1}{C^2(s)}\left(\nabla^2f-\left. \nabla^2f\right|_{\phi=\phi_{A}}\right)\right].
             \label{eq:gamma1}
            \end{aligned}
        \end{equation}
Moreover from (\ref{eq:S2}) the second quantum correction is written as
        \begin{equation}
            \begin{aligned}
            \Gamma_2= & \frac{i \hbar}{2}\left[\left.\int_{0}^{s-\delta s} ds \frac{\operatorname{Vol}^3(X)}{C^3(s)} \nabla^{2}\left(\nabla^{2} f\right)\right|_{\phi=\phi_B}+\int_{0}^{s-\delta s} d s \frac{\operatorname{Vol}^5(X)}{C^5(s)} \left[\nabla\left(\nabla^{2} f\right)\right]^{2}\right|_{\phi=\phi_B} \\ & \left.-\int_{0}^{\bar{s}-\delta s} d s \frac{\operatorname{Vol}^3(x)}{C^3(s)} \nabla^{2}\left(\nabla^{2} f\right)\right|_{\phi=\phi_A}-\left.\int_{0}^{\bar{s}-\delta s} d s \frac{\operatorname{Vol}^5(x)}{C^5(s)} \left[\nabla\left(\nabla^{2} f\right)\right]^{2} \right|_{\phi=\phi_A} \\ & +\int_{\bar{s}-\delta_s}^{\bar{s}+\delta_s} d s \frac{\operatorname{Vol}^3(X)}{C^3(s)}\left(\nabla^{2}\left(\nabla^{2} f\right)-\left.\nabla^{2}\left(\nabla^{2} f\right)\right|_{\phi=\phi_A}\right) \\ & \left.\left.+\int_{\bar{s}-\delta_s}^{\bar{s}+\delta_s} d s \frac{\operatorname{Vol}^5(x)}{C^5(s)}\left(\left[\nabla\left(\nabla^{2} f\right)\right]^{2}-\left[\nabla\left(\nabla^{2} f\right)\right]^{2} \right|_{\phi=\phi_A}\right) \right].
            \end{aligned}
            \label{eq:gamma2}
            \end{equation}
Therefore, with these expressions, we can compute the transition probabilities up to second quantum correction for any model on minisuperspace. Let us remark that we have only assumed a generic form of the Hamiltonian constraint (\ref{HamConst}), thus, these expressions will be valid in principle for any metric and for any gravity theory that leads to a Hamiltonian constraint of this form, in the same way that the semiclassical treatment was valid for Ho\v{r}ava-Lifshitz gravity for example as shown in \cite{Garcia-Compean:2021vcy}. 

However, let us make some assumptions that will be relevant to this article in order to describe some general features of the quantum correction terms. In this work we will only consider General Relativity and the matter content of the system to be a scalar field canonically coupled to gravity with a potential $V(\phi)$ that has a false and a true minimum, thus, the complete action is\footnote{Using natural units in which $c=1$ and $8 \pi G=1$.} 
        \begin{equation}
		S=\frac{1}{2}\int d^4x\sqrt{-g}R-\int d^4x\sqrt{-g}\left[\frac{1}{2}g^{\mu\nu}\partial_{\mu}\phi\partial_{\nu}\phi+V(\phi)\right] .
            \label{eq:ACTION}
        \end{equation} 
This action will lead to a structure of the Hamiltonian constraint which will allow us to recast the function $f$ as
        \begin{equation}
            f=H\left(\varphi^{I}\right)+F\left(\varphi^{I}\right) V(\phi).
        \end{equation}
Furthermore, since the scalar field is canonically coupled, the minisuperspace metric will satisfy $G^{M \phi}=0$ for $M\neq \phi$. Moreover, for the same reason we can assume that $G^{\phi \phi}=G^{\phi \phi}\left(\varphi^{I}\right)$. In addition, as we pointed out earlier, due to the fact that all fields in minisuperspace can be related through eq. (\ref{Relations}), in the regions where the scalar field is constant we can always write  $f=$ $f\left(\Phi^{i}\right)$ for some fixed $i$. In this way, the general solutions (\ref{eq:Cs2}) can be cast as
        \begin{equation}\label{GeneralSolutionsP}
            C(s)= \pm i \operatorname{Vol}(X) \sqrt{\frac{2(\nabla f)^{2}}{f}}, \quad d s=\frac{\operatorname{Vol}(X)(\nabla f)^{2}}{f \nabla^{i} f} d \Phi^{i},
        \end{equation}
then, we can change the integrals in $s$ in the region of constant scalar field in (\ref{eq:gammazero}), (\ref{eq:gamma1}) and (\ref{eq:gamma2}) to integrals in the chosen degree of freedom $\Phi^{i}$.

Moreover, for the third term in (\ref{eq:gammazero}) we note that on the region of the path where the scalar field is not a constant, we have integrals of the general form
	\begin{equation}
		\int_{\bar{s}-\delta s}^{\bar{s}+\delta s}[K(\phi)L(\varphi^{I})]ds ,
	\end{equation}
for some functions $K(\phi)$ and $L(\varphi^{I})$. When dealing with the FLRW metric this term coincides with the tension term in the bubble as found in \cite{Cespedes:2020xpn}. Moreover, in general, we have dealt with these types of terms by defining constants (that we will call tension terms \footnote{Although we will call them \textit{tension} terms, they will not be in general related to the actual tension apart from the term appearing in the semiclassical contribution. For the quantum corrections, they will be independent parameters that should be related to the quantum fluctuations of the scalar field on the wall.}) accompanying the function that depends on the degrees of freedom of the metric evaluated at $\bar{s}$, as previously done in  \cite{Garcia-Compean:2021syl,Garcia-Compean:2021vcy,Garcia-Compean:2022ysy}. This will be valid only on the thin wall limit. Therefore, for the semiclassical contribution we define the tension term as
\begin{equation}
	\left.\operatorname{Vol}(X) F\right|_{\bar{\Phi}^{i}} T_{0}=-2 \operatorname{Vol}(X) i \int_{\bar{s}-\delta s}^{\bar{s}+\delta s} \frac{F}{C(s)}\left(V-V_{A}\right).
	\label{eq:TTerm0}
\end{equation} 
where $\bar{\Phi}^{i}=\Phi^{i}(\bar{s})$. Thus, employing the thin wall limit, we finally obtain that the semiclassical contribution is given by
        \begin{equation}
        \begin{aligned}
            \Gamma_{0}=&\mp \frac{\sqrt{2} {\rm V o l}(X)}{\hbar}\left[\left.\int_{\Phi_{0}^{i}}^{\bar{\Phi}^{i}} \frac{\sqrt{f(\nabla f)^{2}}}{\nabla^{i} f}\right|_{\phi=\phi_{B}} d \Phi^{i}-\left.\int_{\Phi_{0}^{i}}^{\bar{\Phi}^{i}} \frac{\sqrt{f(\nabla f)^{2}}}{\nabla^{i} f}\right|_{\phi=\phi_{A}} d \Phi^{i}\right] \\ & +\left.\frac{{\rm V o l}(X)}{\hbar} F\right|_{\bar{\Phi}^{i}} T_{0},
            \label{eq:gammazeroVOL}
        \end{aligned}
        \end{equation}
where the sign ambiguity on the right hand side appears because of the general solution in (\ref{GeneralSolutionsP}) and will be independent of the sign ambiguity in the left hand side of the expression (\ref{DefGammaGe}). The choice of both signs will be the appropriate choice in order to have well defined probabilities. We note that neither sign ambiguity is physically relevant. The left sign ambiguity just tells us which term is dominant in the solution of the WDW equation and the one on the right hand side tells us which is the appropriate auxiliary function $C(s)$ that we must consider. This expression gives the correct semiclassical results for all the metrics considered in \cite{Garcia-Compean:2021syl}. Furthermore, we note that in general, the transition probability depends on two parameters, in this case, $\bar{\Phi}^{i}$ and $T_{0}$ that we will consider as independent. Moreover, we note that the volume of the spatial slice is an overall constant. Thus, for metrics that do not have a finite value for this volume, we can always compactify the spatial slice and consider appropriate constant values. 

Following the same procedure, for the first quantum correction we obtain in the thin wall limit
        \begin{equation}
            \begin{aligned}
            \Gamma_{1}= &-\frac{\operatorname{Vol}(X)}{2}\left[\left.\int_{\Phi_{0}^{i}}^{\bar{\Phi}^{i}} \frac{\nabla^{2} f}{\nabla^{i} f}\right|_{\phi=\phi_{B}} d \Phi^{i}-\left.\int_{\Phi_{0}^{i}}^{\bar{\Phi}^{i}} \frac{\nabla^{2} f}{\nabla^{i} f}\right|_{\phi=\phi_{A}} d \Phi^{i}\right] \\
            & +\operatorname{Vol}(X)\left[\left.\nabla_{R}^{2} F\right|_{\bar{\Phi}^{i}} T_{1,1}+\left.F G^{\phi \phi}\right|_{\bar{\Phi}^{i}} T_{1,2}\right],
            \end{aligned}
            \label{eq:gamma1VOL}
        \end{equation}
where $T_{1,1}$ and $T_{1,2}$ are tension terms defined analogously as in (\ref{eq:TTerm0})  by
	\begin{equation}\label{eq:TTerm11}
		\left.\operatorname{Vol}(X) \nabla_{R}^{2} F\right|_{\bar{\Phi}^{i}} T_{1,1}= \operatorname{Vol^2}(X)\int_{\bar{s}-\delta s}^{\bar{s}+\delta s} \frac{ds}{C^2(s)}\nabla_{R}^{2} F\left(V-V_{A}\right) ,
	\end{equation}
	\begin{equation}\label{eq:TTerm12}
		\left.\operatorname{Vol}(X) F G^{\phi \phi}\right|_{\bar{\Phi}^{i}} T_{1,2}=\operatorname{Vol^2}(X)\int_{\bar{s}-\delta s}^{\bar{s}+\delta s} \frac{ds}{C^2(s)}F G^{\phi \phi}\left(V^{(2)}-V^{(2)}_{A}\right) ,
	\end{equation}
where $V^{(n)}$ denotes the $n$-th derivative of the potential with respect to the scalar field. Furthermore, we have denoted the restricted Laplacian as
	\begin{equation}
		\nabla_{R}^{2} F=G^{I J} \frac{\partial^{2}}{\partial \varphi^{I} \partial \varphi^{J}} F ,
	\end{equation}
i.e. the Laplacian restricted only to the fields that are not the scalar field. We note that if $\nabla_{R}^{2} F=\zeta F G^{\phi \phi}$ for some constant $\zeta$, or one of these functions is zero, the tension terms can be reduced to only one term. Therefore, we note that in general the first quantum correction will depend on three parameters, one coming from the metric, namely $\bar{\Phi}^{i}$ and two tension terms at most. Furthermore, we note that once again we obtain the global term $\operatorname{Vol}(X)$, thus, there is not any problem with metrics that contain non-compact spatial slices as well. 
 
Finally, following the same procedure we obtain for the  second quantum correction
        \begin{equation}
            \begin{aligned}
            & \Gamma_{2}=\mp \frac{{\rm V o l}(X) \hbar}{4 \sqrt{2}}\left[\left.\int_{\Phi_{0}^{i}}^{\bar{\Phi}^{i}} \frac{\nabla^{2}\left(\nabla^{2} f\right)}{\nabla^{i} f} \sqrt{\frac{f}{(\nabla f)^{2}}}\right|_{\phi=\phi_{B}} d \Phi^{i}-\left.\int_{\Phi_{0}^{i}}^{\bar{\Phi}^{i}} \frac{\nabla^{2}\left(\nabla^{2} f\right)}{\nabla^{i} f} \sqrt{\frac{f}{(\nabla f)^{2}}}\right|_{\phi=\phi_{A}} d \Phi^{i}\right] \\
            & \pm \frac{{\rm V o l}(X) \hbar}{8 \sqrt{2}}\left[\left.\int_{\Phi_{0}^{i}}^{\bar{\Phi}^{i}} \frac{\left[\nabla\left(\nabla^{2} f\right)\right]^{2}}{\nabla^{i} f} \left[\frac{f}{(\nabla f)^{2}}\right]^{3/2}\right|_{\phi=\phi_{B}} d \Phi^{i}-\left.\int_{\Phi_{0}^{i}}^{\bar{\Phi}^{i}} \frac{\left[\nabla\left(\nabla^{2} f\right)\right]^{2}}{\nabla^{i} f} \left[\frac{f}{(\nabla f)^{2}}\right]^{3/2}\right|_{\phi=\phi_{A}} d \Phi^{i}\right] \\
             & +\operatorname{Vol}(X) \hbar\left\{\left.\nabla_{R}^{2}\left(\nabla_{R}^{2} F\right)\right|_{\bar{\Phi}^{i}} T_{2,1}+\left.\left[\left(\nabla_{R}^{2} F\right) G^{\phi \phi}+\nabla_{R}^{2}\left(F G^{\phi \phi}\right)\right]\right|_{\bar{\Phi}^{i}} T_{2,2}+\left.F\left(G^{\phi \phi}\right)^{2}\right|_{\bar{\Phi}^{i}} T_{2,3}\right. \\
            & +\left.2 \nabla_{R}\left(\nabla_{R}^{2} H\right) \cdot \nabla_{R}\left(\nabla_{R}^{2} F\right)\right|_{\bar{\Phi}^{i}} T_{2,4}+\left.\left[\nabla\left(\nabla_{R}^{2} F\right)\right]^{2}\right|_{\bar{\Phi}^{i}} T_{2,5}+\left.G^{\phi \phi}\left(\nabla_{R}^{2} F\right)^{2}\right|_{\bar{\Phi}^{i}} T_{2,6} \\
            & +\left.2 F\left(G^{\phi \phi}\right)^{2} \nabla_{R}^{2} F\right|_{\bar{\Phi}^{i}} T_{2,7}+\left.\nabla_{R}\left(\nabla_{R}^{2} F\right) \cdot \nabla_{R}\left(F G^{\phi \phi}\right)\right|_{\bar{\Phi}^{i}} T_{2,8} \\
            & \left.+\left.2 \nabla_{R}\left(\nabla_{R}^{2} H\right) \cdot \nabla_{R}\left(F G^{\phi \phi}\right)\right|_{\bar{\Phi}^{i}} T_{2,9}+\left.\left[\nabla_{R}\left(F G^{\phi \phi}\right)\right]^{2}\right|_{\bar{\Phi}^{i}} T_{2,10}+\left.F^{2}\left(G^{\phi \phi}\right)^{3}\right|_{\bar{\Phi}^{i}} T_{2,11}\right\} ,
            \end{aligned}
            \label{eq:gamma2VOL}
        \end{equation}
where we have defined the tension terms $T_{2 , i}$ with $i=1,...,11$ in the same general form as above in (\ref{eq:TTerm0}), (\ref{eq:TTerm11}) and (\ref{eq:TTerm12}) with their corresponding functions of the scalar field,  and the sign ambiguity is the same as the one in eq. (\ref{eq:gammazeroVOL}) for both terms. Furthermore, we have defined the dot product in the superspace as
	\begin{equation}
		\quad \nabla f \cdot \nabla g=G^{M N} \frac{\partial f}{\partial \Phi^{M}} \frac{\partial g}{\partial \Phi^{N}}.
	\end{equation}

Therefore, we note that the second quantum correction is described by $\bar{\Phi}^{i}$ one more time but we have many new tension terms, at most we can have $11$ new independent constants (this is the maximum value, it can be reduced if some terms are zero or equivalent). Furthermore, we obtain once again that the spatial slice volume $\operatorname{Vol}(X)$ appears as an overall term. Thus, we can compute the transition probabilities up to second quantum correction for metrics with a non-compact spatial slice.

Let us remark that the number of tension terms increases when the quantum corrections are taken into account. However, as we will see, at least for the metrics that will be of interest to us in the present article, most of the terms will be related or will vanish in such a way that in almost all cases we will only have one tension term per order in the WKB expansion.

Furthermore, let us also point out that in the Euclidean approach the loop contributions are taken into account by writing
	\begin{equation}
		\Phi^{i}=\Phi^{i}_{Cl}+\Phi^{i}_{Q} ,
	\end{equation}
where $\Phi^{i}_{Cl}$ fulfills the equations of motion and $\Phi^{i}_{Q}$ are the quantum fluctuations. In the Lorentzian formalism the results of the quantum corrections are described with the same variable $\bar{\Phi}^{i}$ and employing such decomposition would correspond to the addition of more independent fields to the minisuperspace, which could result in the necessity to expand this formalism to midisuperspace models, which is an intermediate framework with an infinite number of degrees of freedom of the metric but  does not require the full minisuperspace approximation. The quantum corrections that we are computing correspond to considering higher-order terms in the semiclassical WKB expansion, thus, they are quantum corrections of the WDW equation. However, even if we do not employ such a decomposition we can understand these corrections as follows: the quantum fluctuations of the degrees of freedom of the three metric are responsible for adding all the integrals to $\Gamma_{1}$ and $\Gamma_{2}$. Furthermore, the quantum fluctuations of the scalar field on the thin wall are the ones responsible for creating all these new tension terms. Thus, these quantum corrections should correspond to loop corrections in the Euclidean formalism as was proven at one loop for field theory in \cite{deAlwis:2023gth}. Consequently, we can also implement quantum corrections by considering a quantum corrected action. In this form, the Hamiltonian will be modified by adding quantum correction terms to the minisuperspace metric and the $f$ function as in \cite{deAlwis:2023gth}. However, since the resulting Hamiltonian has the same structure as the one studied in the present article, the general formalism presented can be applied to this action as well. Thus, this formalism allows us to pursue quantum correction terms in the form of a quantum corrected action and higher order in the WKB expansion at the same time. However, this is beyond the scope of this work.
\section{Transitions for the closed FLRW metric}
\label{S-PFLRW}
Now that we have described in detail the general method to compute the transition probabilities, let us apply this method for metrics of cosmological interest. Let us begin with the FLRW closed metric that describes a closed homogeneous isotropic universe with positive spatial curvature.

The FLRW metric with positive curvature in 3+1 dimensions is written as
	\begin{equation}
		ds^2=-N^2(t)dt^2+a^2(t)\left(dr^2+\sin^2rd\Omega^2_{2}
		\right) ,
	\end{equation}
where $r\in[0,\pi]$, $d\Omega^2_{2}$ the metric of the sphere, $a(t)$ is the scale factor and $N(t)$ is the lapse function. Considering a homogeneous scalar field we obtain the Hamiltonian constraint
	\begin{equation}
		H=N\left[\frac{\pi_{\phi}^{2}}{2 a^{3}}-\frac{\pi_{a}^{2}}{12 a}-3 a+a^{3} V\right] \simeq 0,
	\end{equation} 
where $V(\phi)$ is the scalar field potential. As always, since the canonical momentum with respect to $N$ vanishes, we can ignore the prefactor and focus only on the term inside brackets in the last expression. Then, by comparing it with the general form (\ref{HamConst}), we identify for this metric a minisuperspace defined by the coordinates $\left\{\Phi^{M}\right\}=\{a, \phi\}$, with inverse metric
       \begin{equation}
                (G^{M N})=\left(\begin{array}{cc}
                                -1 / 6 a & 0 \\
                                0 & 1 / a^3
        \end{array}\right)
        \end{equation}
and
        \begin{equation}
                f(a, \phi)=-3 a+a^{3} V(\phi).
        \end{equation}
We also obtain the volume of the spatial slice as
        \begin{equation}
                \operatorname{Vol}(X)=\int_{\phi=0}^{2 \pi} \int_{\theta=0}^{\pi} \int_{r=0}^{\pi} \sin ^{2} r \sin \theta d r d \theta d \phi=2 \pi^{2}
        \end{equation}
which is finite. In this case, following the general method we can only choose $\Phi^{i}=a$ since there is only one degree of freedom coming from the metric, furthermore, we can choose $a_{0}=a(s=0)=0$ since the expressions are well behaved for this value. Thus, in this case the semiclassical contribution to the transition probability (\ref{eq:gammazeroVOL}) takes the form
\begin{equation}
        \begin{aligned}
            \Gamma_0=&\pm \frac{12 \pi^{2}}{\hbar}\left\{\frac{1}{V_{B}}\left[\left(1-\frac{V_{B}}{3} \bar{a}^{2}\right)^{3 / 2}-1\right]-\frac{1}{V_{A}}\left[\left(1-\frac{V_{A}}{3} \bar{a}^{2}\right)^{3 / 2}-1\right]\right\}\\&+\frac{2 \pi^{2}}{\hbar} \bar{a}^{3} T_{0} ,
        \end{aligned}
        \label{eq:gammazeroPFLRW}
\end{equation}
with $\bar{a}=a(\bar{s})$. This result was derived in \cite{Cespedes:2020xpn,Garcia-Compean:2021syl} and coincides with the one obtained with the Euclidean approach. We note from this result that $\lim_{\bar{a}\to0}\Gamma_{0}=0$, then, taking only this semiclassical contribution we will obtain that the transition probability (\ref{eq:Prob}) fulfills $\lim_{\bar{a}\to0}P(A\to B)=1$. Therefore, interpreting the transition probabilities as probability distributions of creating universes with a given size $\bar{a}$ as it has been previously done in \cite{Cespedes:2020xpn,Garcia-Compean:2021syl,Garcia-Compean:2021vcy,Garcia-Compean:2022ysy}, we note that the semiclassical contribution implies that the most possible scenario is that the universe is created at the spatial singularity.
 
Let us continue with the first quantum correction, that is the term of order $\mathcal{O}\left(\hbar\right)$ in the WKB expansion. The general expression for this term is written in (\ref{eq:gamma1VOL}). For this metric we find that 
        \begin{equation}
               \nabla^2 f = V^{(2)}-V,
        \end{equation}
Therefore, substituting in  (\ref{eq:gamma1VOL}) we obtain
        \begin{equation}
        	\Gamma_{1}=\pi^{2}\left[\left(\frac{V_{B}^{(2)}}{V_{B}}-1\right) \ln \left(1-\bar{a}^{2} V_{B}\right)-\left(\frac{V_{A}^{(2)}}{V_{A}}-1\right) \ln \left(1-\bar{a}^{2} V_{A}\right)\right]+2 \pi^{2} T_{1}.
        	\label{eq:gamma01PFLRW}
        \end{equation}
We note that in this case there is only one tension term since both expressions for the tension terms in the general form are equivalent, thus, the first quantum correction only adds one independent parameter to the transition probability. Moreover we note that $\lim_{\bar{a}\to0}\Gamma_{1}=2\pi^2T_{1}$ which is a constant. Therefore, the inclusion of the first quantum correction will not alter the fact that the point with the biggest probability is the spatial singularity ($\bar{a}=0$) but it reduces the probability in that point.
        
Let us now move on to the second quantum correction. In this case we obtain
        \begin{equation}
            \begin{aligned}
                \nabla^{2}\left(\nabla^{2} f\right) & =\frac{1}{a^3}\left(V^{(4)}-V^{(2)}\right), \\
                \left[\nabla\left(\nabla^{2} f\right)\right]^{2} & = \frac{1}{a^3}\left(V^{(3)}-V^{(1)}\right)^2 .
            \end{aligned}
        \end{equation}
Then, for this metric we have
	\begin{equation}
			\frac{{\rm V o l}(X) \hbar}{4 \sqrt{2}}\left.\int_{\Phi_{0}^{i}}^{\bar{\Phi}^{i}} \frac{\nabla^{2}\left(\nabla^{2} f\right)}{\nabla^{i} f} \sqrt{\frac{f}{(\nabla f)^{2}}}\right|_{\phi=\phi_{B}} d \Phi^{i}= \pi^{2} \hbar\left(V^{(4)}-V^{(2)}\right) F(V_{B}, a)\bigg\rvert^{\bar{a}}_{a_{0}} ,
	\end{equation}
	\begin{equation}
			\frac{{\rm V o l}(X) \hbar}{8 \sqrt{2}}\left.\int_{\Phi_{0}^{i}}^{\bar{\Phi}^{i}} \frac{\left[\nabla\left(\nabla^{2} f\right)\right]^{2}}{\nabla^{i} f} \sqrt{\left(\frac{f}{(\nabla f)^{2}}\right)^{3}}\right|_{\phi=\phi_{B}}=\pi^{2} \hbar (V^{(3)})^2 G(V_{B}, a)\bigg\rvert^{\bar{a}}_{a_{0}} ,
	\end{equation}
where we have defined the functions 
		\begin{equation}
			\begin{aligned}
				F(V, a)=\int \frac{\left(1-\frac{V}{3} \bar{a}^{2}\right)^{1 / 2}}{a\left(1-V \bar{a}^{2}\right)^{2}} d a=-\frac{\sqrt{1-\frac{V}{3} a^{2}}}{2\left(-1+V a^{2}\right)}+\frac{5}{2 \sqrt{6}} \operatorname{arctanh}\left(\frac{\sqrt{3-V a^{2}}}{\sqrt{2}}\right) \\
				-\operatorname{arctanh}\left[\sqrt{1-\frac{V}{3} a^{2}}\right] ,
			\end{aligned}
		\end{equation}
        \begin{equation}
            \begin{array}{r}
                G(V, a)=\int \frac{a\left(1-\frac{V}{3} \bar{a}^{2}\right)^{3 / 2}}{\left(1-V \bar{a}^{2}\right)^{4}} d a=\frac{1}{576 \sqrt{3} V}\left[-\frac{2 \sqrt{3-V a^{2}}\left(63-34 V a^{2}+3 V^{2} a^{4}\right)}{\left(-1+V a^{2}\right)^{3}}\right. \\
                \left.-3 \sqrt{2} \operatorname{arctanh}\left(\frac{\sqrt{3-V a^{2}}}{\sqrt{2}}\right)\right].
            \end{array}   
        \end{equation}
We note that the function $G(V, a)$ is well behaved in the limit $a\to0$. On the other hand, the third term of the $F(V,a)$ function diverges in this limit. However, we note that this limit is independent of $V$, therefore  we can eliminate this divergence in the transition probability by imposing the condition
        \begin{equation}\label{eq:ConditionFLRW}
            V_{B}^{(4)}-V_{B}^{(2)}=V_{A}^{(4)}-V_{A}^{(2)}.
        \end{equation}
In this way, we can still choose  $a_{0}=0$ in order to have access to the initial spatial singularity. Furthermore, for this metric from the eleven possible tension terms, only one remains since all the non-zero terms have the same dependence on $\bar{a}$, thus, we finally obtain that the second quantum correction is written as
        \begin{equation}
            \begin{aligned}
                \Gamma_{2}= &\pm \pi^{2} \hbar\left\{\left(V_{B}^{(4)}-V_{B}^{(2)}\right)\left[F\left(V_{B}, \bar{a}\right)-F\left(V_{A}, \bar{a}\right)\right]-\right. \left(V_{B}^{(3)}\right)^{2}\left[G\left(V_{B}, \bar{a}\right)-G\left(V_{B}, 0\right)\right] \\
                & \left.+\left(V_{A}^{(3)}\right)^{2}\left[G\left(V_{A}, \bar{a}\right)-G\left(V_{A}, 0\right)\right]\right\}+\frac{2\pi^{2}\hbar }{\bar{a}^{3}} T_{2}.
            \end{aligned}
            \label{eq:gamma012PFLRW}
        \end{equation}
We note that once again, the second quantum correction only adds one more independent parameter tension term. From this result we also note that the limit $\lim_{\bar{a}\to0}\Gamma_{2}$ will diverge because of the tension term. Thus, choosing appropriately the sign of (\ref{DefGammaGe}) we will obtain that taking into account up to second order in the quantum corrections the transition probability fulfills $\lim_{\bar{a}\to0}P(A\to B)=0$. Therefore, the second order quantum correction implies that the most probable size of the universe to be created is at a finite non-zero value of the scale factor. In this way, the quantum correction terms lead to a result that avoids the spatial singularity. 

Let us remark from (\ref{eq:gammazeroPFLRW}) that the semiclassical contribution $\Gamma_{0}$ is well behaved for all values of the potentials, but it can be complex for positive values of the potential minima. However, this is not a problem since we always take only the real part. Nevertheless, we see from (\ref{eq:gamma01PFLRW}) that $\Gamma_{1}$ has a divergence in $\bar{a}^{2}=V_{A, B}^{-1}$ which can occur for positive values of the potential minima. Furthermore from (\ref{eq:gamma012PFLRW}) we note that $\Gamma_{2}$ will also have a divergence appearing in $\bar{a}^{2}=\frac{5}{V_{A, B}}$, which can also appear for positive values of the potential minima. In these cases, the transition probability will only be well defined up to an upper limit. On the other hand, we note that for negative potential minima everything is well defined without any issue. Therefore positive values of the potential minima will imply some additional restrictions for the validity of the transition probabilities. Moreover, in order to have a well defined second quantum correction term we need to impose additional restrictions on the higher derivatives of the potential expressed in (\ref{eq:ConditionFLRW}).

\begin{figure}[ht]
            \centering
            \includegraphics[width=\textwidth]{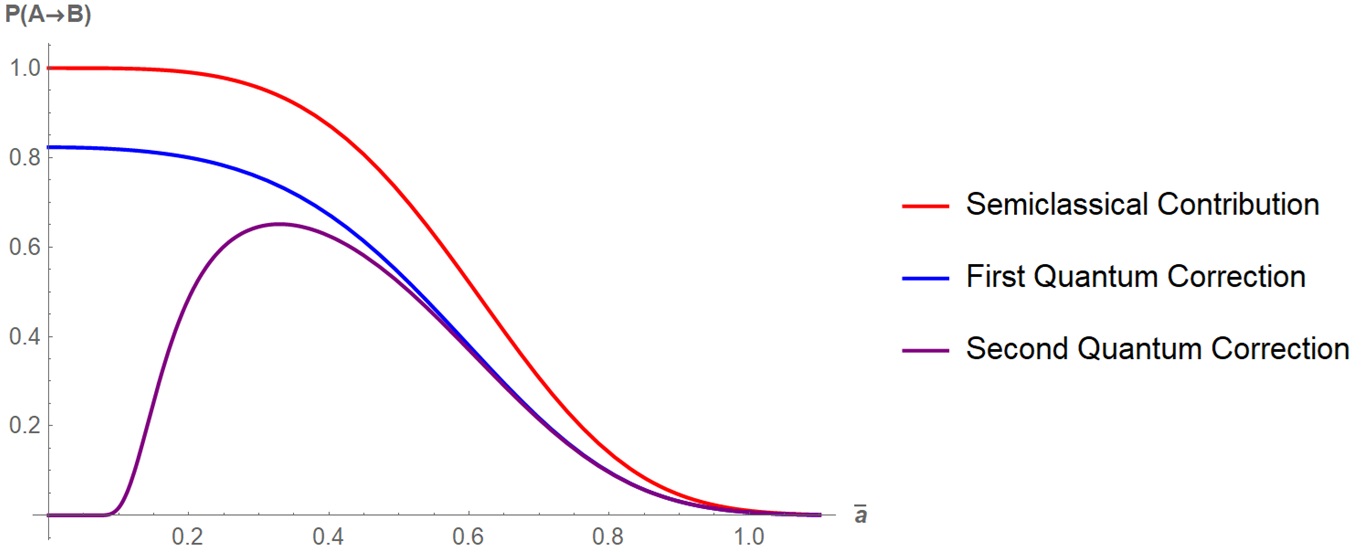}
                \caption{Transition probability for the positive FLRW metric in units such that $\hbar=1$, choosing $V_{B}=-1.5$, $V_{A}=-1$, $\frac{V_{B}^{(2)}}{V_{B}}-1=\frac{V_{A}^{(2)}}{V_{A}}-1=0.05$, $V_{B}^{(4)}-V_{B}^{(2)}=-0.005$, $V_{B}^{(3)}=V_{A}^{(3)}=0.1$, $T_{0}=0.005$, $T_{1}=5\times10^{-4}$, $T_{2}=10^{-6}$. We plot the results with the semiclassical contribution $\Gamma_0$ (red line), including the  first quantum correction $\Gamma_0+\Gamma_1$ (blue line) and including the second quantum correction $\Gamma_0+\Gamma_1+\Gamma_2$ (purple line).}
            \label{fig:POSITIVEFLRWWC}
        \end{figure}

In order to explicitly visualize the effect of the first and second order quantum corrections for all values of the scale factor, we plot in Figure \ref{fig:POSITIVEFLRWWC} the transition probability (\ref{eq:Prob}) using the results for this metric, that is using eq. (\ref{eq:gammazeroPFLRW}) for the semiclassical contribution, eq. (\ref{eq:gamma01PFLRW}) for the first quantum correction and eq. (\ref{eq:gamma012PFLRW}) for the second order quantum correction. We choose units in which $\hbar=1$, we also choose $V_{B}=-1.5$, $V_{A}=-1$, $\frac{V_{B}^{(2)}}{V_{B}}-1=\frac{V_{A}^{(2)}}{V_{A}}-1=0.05$, $V_{B}^{(4)}-V_{B}^{(2)}=-0.005$, $V_{B}^{(3)}=V_{A}^{(3)}=0.1$, $T_{0}=0.005$, $T_{1}=5\times10^{-4}$ and $T_{2}=10^{-6}$. In order to have a well defined transition probability we choose the plus sign in the left hand side of (\ref{DefGammaGe}) and the minus sign in the right of (\ref{eq:gammazeroPFLRW}) which corresponds to the minus sign in (\ref{eq:gamma012PFLRW}) as well.  We note that the result with the semiclassical contribution only has a maximum value on the spatial singularity and as $\bar{a}$ increases, the probability decreases. The result including the first quantum correction does not change the overall shape of the curve, it only reduces the overall probability and makes it fall faster. However, as we have previously seen, when the second quantum correction is taken into account as well, the shape of the curve is changed in the ultraviolet (small values of the scale factor), in such a way that the maximum value of the transition probability is no longer in the spatial singularity, it is instead located at a small non-zero value of the scale factor. From this point, the same behaviour is encountered as in the other two results in the infrared, that is as the scale factor increases, the probability decreases. Thus, the quantum corrections are only relevant on the ultraviolet as expected. We remark that this behaviour is independent of the specific numerical values of the parameters, we only require that they have the same signs as the particular choices that we employed.

\section{Transitions for the flat FLRW metric}
\label{S-FFLRW}
In this section we consider the FLRW metric with zero spatial curvature. This metric will be important in the following since it represents the isotropic limit of the Bianchi III metric. We can write this metric in Cartesian coordinates as follows
        \begin{equation}
                d s^{2}=-N(t) d t^{2}+a^{2}(t)\left[d x^{2}+d y^{2}+d z^{2}\right].
        \end{equation}
The Hamiltonian constraint is given by
        \begin{equation}
            H=N\left[\frac{\pi_{\phi}^{2}}{2 a^{3}}-\frac{\pi_{a}^{2}}{12 a}+a^{3} V\right] \simeq 0.
            \label{eq:HamFFLRW}
        \end{equation}
As always, the $N$ function is not dynamical, so we can focus on the terms within brackets. Comparing it with the general form (\ref{eq:WDWGR}) we see that once again the minisuperspace is defined by the coordinates $\left\{\Phi^{M}\right\}=\{a, \phi\}$ with the metric
        \begin{equation}
            (G^{M N})=\left(\begin{array}{cc}
            -\frac{1}{6 a} & 0 \\
            0 & \frac{1}{a^3} 
            \end{array}\right)
            \label{eq:GMNFlatNF}
        \end{equation}
and
        \begin{equation}
             f(a, \phi)=a^{3} V(\phi).
        \end{equation}
We also note that in this case the spatial volume takes the form
        \begin{equation}
            \operatorname{Vol}(X)=\iiint d x d y d z.
        \end{equation}
In a strict sense, this volume is divergent as a result of the non-compact nature of the 3-space of the 3+1 decomposition. However, as we pointed out earlier we can constrain the variables to a finite interval in order to obtain a finite value for this term. Since it will be an overall constant we will not worry about its value and we will maintain it as an arbitrary constant that will not modify the behaviour of the transition probabilities.

Then, substituting the above identifications for this metric into eq. (\ref{eq:gammazeroVOL}) we obtain
        \begin{equation}
            \Gamma_0= \pm \frac{2 i \operatorname{Vol}(X)}{\sqrt{3} \hbar}\left(\sqrt{V_{B}}-\sqrt{V_{A}}\right) \bar{a}^{3}+\frac{\operatorname{Vol}(X)}{\hbar} \bar{a}^{3} T_{0}.
           \label{eq:G0FLRWF}
        \end{equation}
We note that $\lim_{\bar{a}\to0}\Gamma_{0}=0$, then, considering only the semiclassical contribution we would obtain for the transition probability $\lim_{\alpha=0}P(A\to B)=1$. Thus, once again, the semiclassical contribution leads to a maximum value for probability at the initial singularity.

Proceeding with the first quantum correction, we have also in this case 
        \begin{equation}
             \nabla^2 f=V^{\prime \prime}-V,
             \label{LaplacianFLRW}
        \end{equation}
that substituting back into (\ref{eq:gamma1VOL}) we get
        \begin{equation}
         \Gamma_1=\left.\frac{\operatorname{Vol}(X)}{2}\left(\frac{V_B^{(2)}}{V_B}-\frac{V_A^{(2)}}{V_A}\right) \ln(a)\right|_{a_0} ^{\bar{a}} +\frac{\operatorname{Vol}(X)}{2}T_{1}.
         \label{eq:G1FLRFAux}
        \end{equation}
Moreover, as in the previous case, there only remains a single tension term, thus, the first quantum correction only adds one independent constant. From this expression we also notice that the choice $a_0=0$ may lead to a divergence due to the logarithmic term. Therefore, in order to obtain a consistent probability we are forced to impose the condition on the second derivatives of the potential
	\begin{equation}
		\frac{V_B^{(2)}}{V_B}=\frac{V_A^{(2)}}{V_A}.
		\label{eq:ConditionsFirstQC}
	\end{equation}
Thus, the first quantum correction is written finally as
	\begin{equation}
		\Gamma_1=\frac{\operatorname{Vol}(x)}{2}T_{1} ,
		\label{eq:G1FLRF}
	\end{equation}
which is just a constant. Consequently, as was discussed in the previous example, the first quantum correction does not alter the overall behaviour, in particular the point with maximum probability.

Let us move on to the computation of the second quantum correction. For this metric we also have 
        \begin{equation}
            \nabla^{2}\left(\nabla^{2} f\right)= \frac{V^{(4)}-V^{(2)}}{a^3} ,
        \end{equation}
         \begin{equation}
            \left[\nabla\left(\nabla^{2} f\right)\right]^{2}=\frac{\left( V^{(3)}-V^{(1)} \right)^2}{a^3}.
        \end{equation}
Therefore, substituting back these equations into (\ref{eq:gamma2VOL}) and computing all the remaining terms, we finally obtain
        \begin{equation}
            \begin{aligned}
            \Gamma_2= &\pm\frac{i \hbar \operatorname{Vol}(X)}{2}\left[\left.\left(W_B-W_A\right) \frac{1}{3a^3 }\right|_{a_0} ^{\bar{a}}\right] + \frac{\hbar\operatorname{Vol}(X)}{ \bar{a}^3}T_{2},
            \label{eq:G2FLRWF}
            \end{aligned}
        \end{equation}
where we have defined 
	\begin{equation}
		W=\frac{V^{(2)}-V^{(4)}}{\sqrt{3}V^{3 / 2}}-\frac{\left(V^{(3)}\right)^2}{3\sqrt{3}V^{5/2}} .
	\end{equation}
For this metric we obtain once again that from the eleven possible different tension terms, only one dependence on $\bar{a}$ appears, leading to only one tension term. In this way, the second quantum correction also adds only one independent constant. Here we also note that the choice $a_{0}=0$ will lead to a divergence in the first term of (\ref{eq:G2FLRWF}). However, for positive values of the potential minima we note that the first term never contributes to the transition probability after taking the real part. Therefore, for positive potentials the second quantum correction takes simply the form
	\begin{equation}
		\Gamma_2=  \frac{\operatorname{Vol}(X)\hbar}{ \bar{a}^3}T_{2} .
		\label{eq:G2FLRWFF}
	\end{equation}
On the other hand, for negative potential minima, this divergence can be eliminated by imposing the constraint
	\begin{equation} \label{eq:ConditionFlatFLRW2}
		W_{A}=W_{B} .
	\end{equation}
Then, the negative potential minima leads to an additional constraint on the higher derivative terms of the potential, but it takes the same form as in the previous example. In any case, we note that $\lim_{\alpha=0}\Gamma_{2}$ will diverge. Thus, by choosing the proper signs, we will obtain that $\lim_{\bar{a} \rightarrow 0}P(A\to B)=0$ as in the previous section. 

We can plot the transition probability for this metric considering the semiclassical contribution (\ref{eq:G0FLRWF}), the first quantum correction (\ref{eq:G1FLRF}) and the second quantum correction (\ref{eq:G2FLRWFF}). In this case, we obtain the same general behaviour as the one encountered for the positive FLRW metric in Figure \ref{fig:POSITIVEFLRWWC} in the previous section. The only difference is that the conditions for the higher derivatives of the potential are modified, in this case we must impose (\ref{eq:ConditionsFirstQC}) and (\ref{eq:ConditionFlatFLRW2}). Therefore, for this metric the second quantum correction will also lead to an avoidance of the initial singularity.

\section{Transitions for the Bianchi III metric}
\label{S-B3}
Now that we have completed the study of the FLRW metrics, let us move on to consider metrics describing homogeneous but anisotropic universes. We will begin by considering the Bianchi III metric in the present section. In a local set of coordinates we can write this metric in the following form \cite{Akarsu:2009gz,VijayaSanthi:2020feh,Ellis:1968vb}
\begin{equation}
	d s^{2} = -N^{2}(t) d t^{2}+A^{2}(t) d x^{2}+B^{2}(t) e^{-2 \alpha x} d y^{2}+C^{2}(t) d z^{2},
	\label{eq:BIII}
\end{equation}
where $\alpha \neq 0$ is a constant measuring the amount of anisotropy. We see that the isotropy limit is performed by taking $A(t)=B(t)=C(t)$ and $\alpha\to0$ and results in the  flat FLRW metric by considering the coordinates to correspond to Cartesian coordinates in this limit. Furthermore, if we consider different values for the scale factors $A(t)$, $B(t)$ and $C(t)$ and take $\alpha\to0$ we will obtain the anisotropic Bianchi I metric.

Considering a homogeneous scalar field coupled to gravity the Hamiltonian constraint takes the form
\begin{equation}
	\begin{aligned}
		H=N&\left[\frac{A}{4 B C} \pi_{A}^{2}+\frac{B}{4 A C} \pi_{B}^{2}+\frac{C}{4 A B} \pi_{C}^{2}-\frac{1}{2 C} \pi_{A} \pi_{B}-\frac{1}{2 B} \pi_{A} \pi_{C}-\frac{1}{2 A} \pi_{B} \pi_{C}\right. \\
		&\left.+\frac{\pi_{\phi}^{2}}{2 A B C}+\frac{\alpha^{2} B C}{A}+V(\phi) A B C\right]\simeq0 .
	\end{aligned}
	\label{eq:HamConstB3}
\end{equation}

Comparing with the general form (\ref{HamConst}), we see that the coordinates on minisuperspace are $\left\{\Phi^{M}\right\}=\{A, B, C, \phi\}$. Moreover the metric is written as
\begin{equation}
	\left(G^{M N}\right)=\frac{1}{2ABC}\left(\begin{array}{cccc}
		A^{2} & -A B & -A C & 0 \\
		-A B & B^{2} & -B C & 0 \\
		-A C & -B C & C^{2} & 0 \\
		0 & 0 & 0 & 2
	\end{array}\right)
\end{equation}
and the function $f$ reads
\begin{equation}
	f(A, B, C, \phi)=ABC\left(V(\phi)+\frac{\alpha^{2}}{A^2}\right).
\end{equation}
We also note that in this case the volume of $X$ is given by
\begin{equation}
	\operatorname{Vol}(X)=\iiint e^{-\alpha x} d x d y x d z
\end{equation}
which again is finite only if we restrict the spatial slice to a finite interval in $y$ and $z$ but we will consider it as an arbitrary constant. In \cite{Garcia-Compean:2021syl} the semiclassical contribution to the transition probability was studied for this metric. However, in that work, a convenient factorization was used to obtain a result consistent in the case of a vanishing potential. In the present case, it will be more convenient to consider the Hamiltonian constraint without any factorization. The dependence of the results on a particular factorization will be explored in Appendix \ref{S-AP}.

We note that for this metric we have more than one variable on the minisuperspace apart from the scalar field. Thus, as it was asserted in the general method in Section \ref{S-KSNormal} these variables will be related as a consequence of the semiclassical expansion. In particular, in the region where the scalar field is constant eq. (\ref{Relations}) leads to
	\begin{equation}
		\frac{dB}{dC}=\frac{B}{C} , \hspace{1cm} \frac{dA}{dB}=\frac{A\left(V+\frac{3\alpha^2}{A^2}\right)}{B\left(B-\frac{\alpha^2}{A^2}\right)} ,
	\end{equation}
which can be integrated to obtain
	\begin{equation}
		B=b_{0}C=b_{1}\frac{\left(A^2+\frac{3\alpha^2}{V}\right)^{2/3}}{A^{1/3}} ,
	\end{equation}
where $b_{0}$ and $b_{1}$ are integration constants. However, we can set these constants to $1$ by rescaling the variables on the metric in a suitable way for each of the solutions and demanding continuity on the wall at $\bar{B}$. Therefore we obtain for the region where the scalar field is constant
	\begin{equation}\label{RelationsBianchi3}
			B=C=\frac{\left(A^2+\frac{3\alpha^2}{V}\right)^{2/3}}{A^{1/3}} .
	\end{equation}
From this relation, we note that the isotropy limit leads to $\lim_{\alpha\to0}B=A$. Therefore, all the anisotropy will be measured by the constant $\alpha$, Moreover, as we pointed out previously the limit $\alpha\to0$ corresponds to the Bianchi I metric, thus, from these relations we note that if we study the Bianchi I metric, we will be forced to obtain the flat FLRW metric and its corresponding result. Thus, the Bianchi III metric is more suitable to study the effect of anisotropy.

Furthermore, we note that given $\alpha\neq0$, the variable $B$ will diverge as $A\to0$. On the other hand for big values of $A$, $B$ increases with $A$. Thus, in this case the cosmological behaviour does not describe a spatial singularity at the beginning ($A=0$). Instead, we have in the early universe (understood as small values of $A$) a behaviour in which $B$ and $C$ start with big values (infinite in principle), then, as $A$ grows, they decrease until they reach a minimum value, from where they start growing. For positive potential minima, this minimum value is non-zero, whereas for a negative potential minimum it is zero but always with a non-zero value of $A$. Thus, we have a phase of contraction on the $B$ and $C$ direction, until they reach a point where they start an expansion as $A$ expands, that is, similar to a bounce scenario as the dimension described by $A$ emerges\footnote{The relation between the coordinates can change if we consider a different factorization. This point will be discussed in the Appendix \ref{S-AP}.}. We note that the size of the contraction phase before the bounce depends on the size of the values of $\frac{3\alpha^2}{V}$ and it disappears in the isotropy limit. Let us remark that since we have two different degrees of freedom coming from the metric, namely $A$ and $B$ we can write down the transition probabilities in terms of any of these two, however only $A$ is single-valued in time, whereas the other two can have the same value twice given the bounce nature. For all these reasons we will consider only positive values of the potential minima and from now on write the transition probabilities in terms of the $A$ function.

With the identifications made above for this metric, substituting back into (\ref{eq:gammazero}) we obtain for the semiclassical contribution
	\begin{equation}
		 \begin{aligned}
		\Gamma_{0}=&\pm\frac{2i {\rm Vol}(X)}{\hbar}\left[\sqrt{V_{B}}F_{III}(V_{B},A)\bigg\rvert_{A_{0}}^{\bar{A}}-\sqrt{V_{A}}F_{III}(V_{A},A)\bigg\rvert_{A_{0}}^{\bar{A}}\right]\\&+\frac{{\rm Vol}(X)}{\hbar}\bar{A}^{1/3}\left(\bar{A}^2+\frac{3\alpha^2}{V_{B}}\right)^{4/3}T_{0} ,
		\label{GammaZeroB3}
	\end{aligned}
	\end{equation}
where we have defined the function
	\begin{equation}
		F_{III}(V,A)=\int\frac{\left(A^2+\frac{3\alpha^2}{V}\right)^{1/3}}{A^{5/3}}\sqrt{\left(A^2+\frac{\alpha^2}{V}\right)\left(A^2-\frac{\alpha^2}{V}\right)\left(3A^2+\frac{5\alpha^2}{V}\right)}dA .
	\end{equation}
It can be shown that the Bianchi III results (\ref{GammaZeroB3}) coincides with the flat FLRW result (\ref{eq:G0FLRWF}) in the isotropy limit $\alpha\to0$ as we expected.

\begin{figure}[ht]
            \centering
            \includegraphics[width=\textwidth]{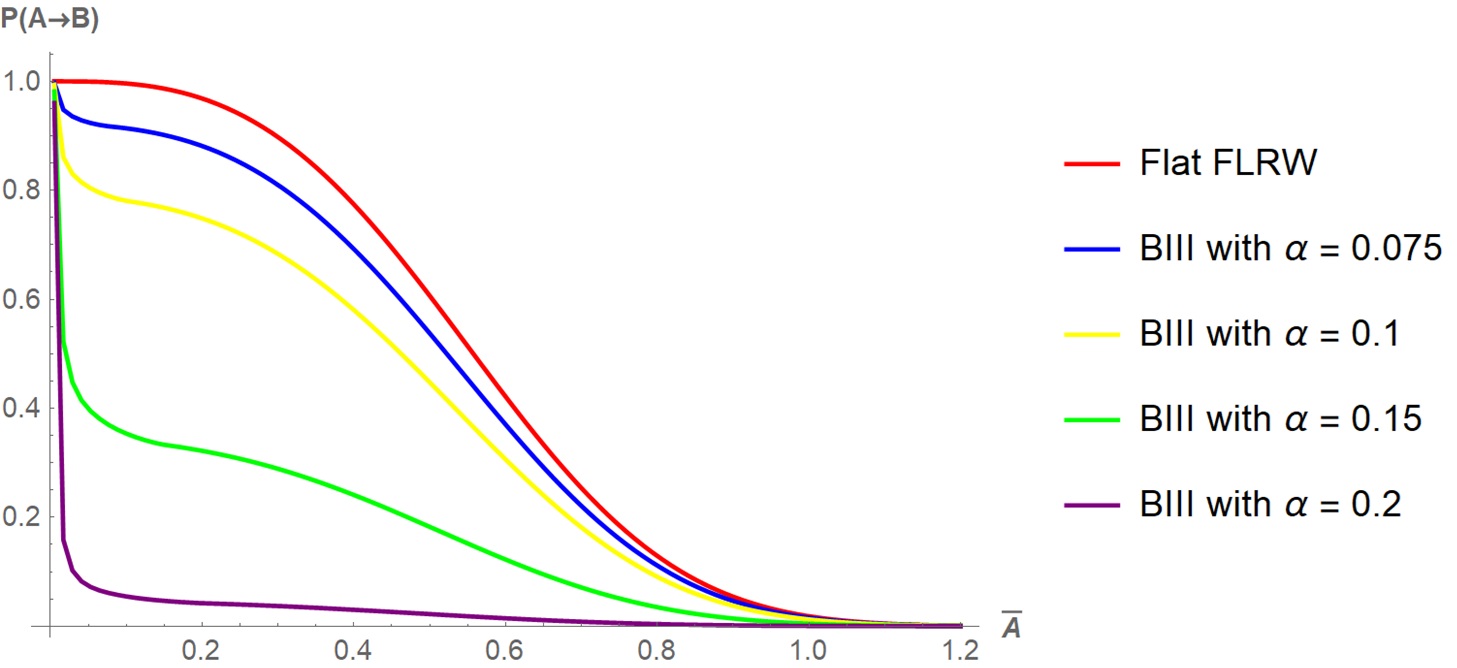}
                \caption{Transition probabilities for the Bianchi III metric in units such that $\hbar=1$, choosing $V_{B}=1$, $V_{A}=5$, $T_{0}=1$. We plot the results of the semiclassical contribution $\Gamma_0$ for different values of $\alpha$. We choose  $\alpha=0$ that corresponds to the flat FLRW case in which $\bar{A}\to\bar{a}$ (red line), $\alpha=0.075$ (blue line), $\alpha=0.1$ (yellow line), $\alpha=0.15$ (green line) and $\alpha=0.2$ (purple line).}
            \label{fig:SCCBIII}
        \end{figure}

Taking $A_{0}=0$ we note from the last expression that $\lim_{\bar{A}\to0}\Gamma_{0}=0$. Thus, we obtain once again that considering only the semiclassical contribution the probability will have its biggest value in $\bar{A}=0$. However, let us remark that in this case this point does not represent a spatial singularity. Furthermore, we plot the last expression for different values of  $\alpha$ to obtain Figure \ref{fig:SCCBIII}, where in order to have a well-behaved probability we have chosen the minus sign in both sign ambiguities. Additionally, we choose positive values for the potential minima, $V_B=1$ and $V_A=5$, and for the tension term $T_{0}=1$. We can see that the form of the semiclassical contribution coincides with the general form of the FLRW results. Moreover, the effect of anisotropy is to decrease the value of the transition probability as the value of $\alpha$ increases, and the case $\alpha\to0$ corresponds to the FLRW result, in this form our result is consistent with \cite{Garcia-Compean:2021syl}. 

\begin{figure}[ht]
            \centering
            \includegraphics[width=\textwidth]{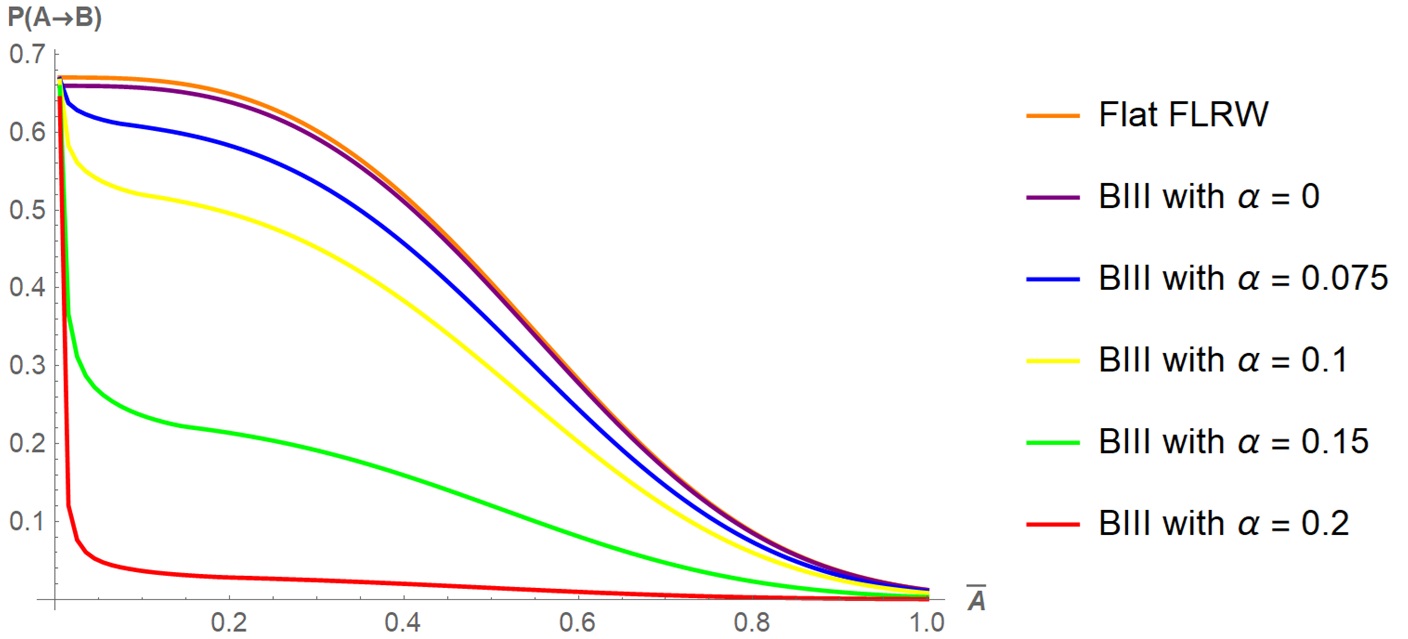}
                \caption{Transition probabilities for the flat FLRW (orange line) and Bianchi III metrics in units such that $\hbar=1$, choosing $V_{B}=1$, $V_{A}=5$, $\frac{V^{(2)}_{B}}{V_{B}}-\frac{11}{3}=-0.005$, $T_{0}=1$, $T_1=0.1$. We plot the results of the semiclassical contribution plus the first quantum correction, $\Gamma_0+\Gamma_1$, taking different values of $\alpha$; we choose the cases $\alpha=0$ (purple line), $\alpha=0.075$ (blue line), $\alpha=0.1$ (yellow line), $\alpha=0.15$ (green line) and $\alpha=0.2$ (red line). The isotropy limit of the Bianchi III metric $\alpha \to 0$, does not correspond exactly to the flat FLRW result.}
            \label{fig:FQCBIII}
        \end{figure}

Let us continue now with the computation of the first quantum correction, in this case we obtain
\begin{equation}
	\nabla^2f=V^{(2)}-3V+\frac{2\alpha^2}{A^2} .
	\label{LaplacianB3}
\end{equation}
Then, computing the remaining terms and substituting back into eq. (\ref{eq:gamma1VOL}) we obtain for the first quantum correction
\begin{equation}
	\begin{aligned}
		\Gamma_1=\operatorname{Vol}(X) & \left[\left(\frac{V^{(2)}_{B}}{V_{B}}-\frac{11}{3}\right)\ln\sqrt{\frac{\bar{A}^2+\frac{3\alpha^2}{V_{B}}}{A^2_{0}+\frac{3\alpha^2}{V_{B}}}}-\left(\frac{V^{(2)}_{A}}{V_{A}}-\frac{11}{3}\right)\ln\sqrt{\frac{\bar{A}^2+\frac{3\alpha^2}{V_{A}}}{A^2_{0}+\frac{3\alpha^2}{V_{A}}}}\right]\\&+{\rm Vol}(X)T_{1}.
	\end{aligned}
	\label{eq:Gamma1B3Aux}
\end{equation}
We note that taking the isotropic limit $\alpha\to0$ in the last expression we obtain the flat FLRW result of eq. (\ref{eq:G1FLRF}). However, contrary to that case, we will not have any divergence if we take $A_{0}=0$ as long as $\alpha\neq0$. Moreover, in order to obtain the flat FLRW result in the limit $\alpha\to0$ we impose the same condition (\ref{eq:ConditionsFirstQC}), thus, the first quantum correction takes the form 
	\begin{equation}
			\Gamma_1=\operatorname{Vol}(X)  \left\{\left(\frac{V^{(2)}_{B}}{V_{B}}-\frac{11}{3}\right)\ln\left[\sqrt{\frac{\bar{A}^2+\frac{3\alpha^2}{V_{B}}}{A^2_{0}+\frac{3\alpha^2}{V_{B}}}}\sqrt{\frac{\bar{A}^2+\frac{3\alpha^2}{V_{A}}}{A^2_{0}+\frac{3\alpha^2}{V_{A}}}}\right]\right\}+ {\rm Vol}(X)T_{1}.
	\end{equation}
Taking $A_{0}=0$, the latter simplifies to
	\begin{equation}
			\Gamma_1=\operatorname{Vol}(X)  \left(\frac{V^{(2)}_{B}}{V_{B}}-\frac{11}{3}\right)\ln\sqrt{\frac{V_{B}}{V_{A}}\left(\frac{\bar{A}^2+\frac{3\alpha^2}{V_{B}}}{\bar{A}^2+\frac{3\alpha^2}{V_{A}}}\right)}+ {\rm Vol}(X)T_{1}.
		\label{eq:Gamma1B3}
	\end{equation}
However, we note that once we have chosen $A_{0}=0$ we can no longer recover the flat FLRW result (\ref{eq:G1FLRF}) in the isotropy limit. In the isotropy limit the first quantum correction will be a constant, but the first term will contribute with a logarithmic constant term as well. Thus, the first quantum correction does not lead exactly to the flat FLRW result in the isotropy limit. Furthermore, we also note that we only obtain one tension term, thus, the first quantum correction adds only one independent constant. Moreover, we note that in order to have a well defined probability we should choose only positive values of the potential minima as we have remarked before. In addition, we also note that $\lim_{\bar{A}\to0}\Gamma_{1}={\rm Vol}(X)T_{1}$ which is a constant, thus, as in the previous cases the first quantum correction will not alter the behaviour near the value $\bar{A}=0$. In Figure \ref{fig:FQCBIII} we show a plot of this result taking different values of $\alpha$, we have considered the same values for the potential minima, and choose $\frac{V^{(2)}_{B}}{V_{B}}-\frac{11}{3}=-0.005$, $T_{0}=1$, $T_1=0.1$. We see that the general behaviour is the same as the FLRW result. That is, for all values of $\alpha$, the first quantum correction does not change the overall behaviour, it only reduces the probability. The effect of anisotropy is to reduce the probability once again. Moreover, in this case, the isotropy limit does not correspond exactly to the flat FLRW result.

Finally, for the second correction we obtain in this case
\begin{equation}
	\nabla^{2}\left(\nabla^{2} f\right)= \frac{V^{(4)}-3V^{(2)}+\frac{6\alpha^2}{A^2}}{ABC} ,
\end{equation}
\begin{equation}
		\left[\nabla\left(\nabla^{2} f\right)\right]^{2}=\frac{(V^{(3)}-3V^{(1)})^2+\frac{8\alpha^4}{A^4}}{ABC} .
\end{equation}

Then, computing the remaining terms in eq. (\ref{eq:gamma2VOL}) we obtain for the second quantum correction
	\begin{equation}
		\begin{aligned}
			\Gamma_2&=\pm\frac{i\hbar\operatorname{Vol}(X)}{2} \left[\frac{1}{V^{3/2}_{B}}H_{III}(V_{B},A)\bigg\rvert_{A_{0}}^{\bar{A}}-\frac{1}{V^{3/2}_{A}}H_{III}(V_{A},A)\bigg\rvert_{A_{0}}^{\bar{A}}\right]\\&+\frac{{\rm Vol}(X)\hbar}{\bar{A}^{1/3}\left(\bar{A}^2+\frac{3\alpha^2}{V_{B}}\right)^{4/3}}T_{2},
		\end{aligned}
            \label{eq:gamma2BIII}
	\end{equation}
where the $H_{III}$ function is defined as
	\begin{equation}
		\begin{aligned}
		H_{III}(V,A)=&\int\frac{dA}{A^{1/3}\left(A^2+\frac{3\alpha^2}{V}\right)^{7/3}}\sqrt{\frac{A^2+\frac{\alpha^2}{V}}{(A^2-\frac{\alpha^2}{V})(3A^2+\frac{5\alpha^2}{V})}}
		\\ &\times \left[6\alpha^2+\left(V^{(4)}-3V^{(2)}\right)A^4+\frac{\left((V^{(3)})^2A^4+8\alpha^2\right)\left(A^2+\frac{\alpha^2}{V}\right)}{V\left(A^2-\frac{\alpha^2}{V}\right)\left(3A^2+\frac{5\alpha^2}{V}\right)}\right] .
		\end{aligned}
            \label{eq:HIII}
	\end{equation}
As in the previous case we note that the first term of eq. (\ref{eq:gamma2BIII}) has an overall $i$, thus, if the function is always real for positive potentials, this term will not contribute. Furthermore, we note that once again only one tension term is added from the eleven possible because the non-zero terms have the same dependence with $A$. However, if we take the isotropy limit for this result we obtain
	\begin{equation}
		\begin{aligned}
			\lim_{\alpha=0}\Gamma_2= &\pm\frac{i \hbar \operatorname{Vol}(X)}{2}\left[\left.\left(\bar{W}_B-\bar{W}_A\right) \frac{1}{3A^3 }\right|_{A_0} ^{\bar{A}}\right] + \frac{\hbar\operatorname{Vol}(X)}{ \bar{A}^3}T_{2},
		\end{aligned}
	\end{equation}
	where
	\begin{equation}
		\bar{W}=\frac{\left(3V^{(2)}-V^{(4)}\right)}{\sqrt{3}V^{3 / 2}}-\frac{\left(V^{(3)}\right)^2}{3\sqrt{3}V^{5/2}} ,
	\end{equation}
which is very similar to the result for the FLRW metric in eq. (\ref{eq:G2FLRWF}). The only difference appears in the constant accompanying the second derivative on the $W$ function, in this case it is a factor of $3$, whereas for the flat FLRW metric it was just a factor of $1$. The differences that appear at the first and the second quantum corrections come from the fact that although the FLRW metric follows from the isotropy limit as a classical result, the metric on superspace for the flat FLRW case does not come from the Bianchi III metric in this limit, thus, the structure of the minisuperspace is not the same after taking the isotropy limit and we should not expect the same results in all scenarios. In particular, we note that the Laplacian in the Bianchi III  metric (\ref{LaplacianB3}) does not lead to the Laplacian for the FLRW metric (\ref{LaplacianB3}) in the isotropy limit. It is interesting that this discrepancy appears when we take into account the quantum corrections and it only changes by constants, it does not change the general behaviour. Furthermore, we note that in the general case we have for this metric that $\lim_{\bar{A}\to0}\Gamma_{2}$ will diverge, thus, making the probability vanish on this limit after choosing the appropriate signs.

\begin{figure}[ht]
            \centering
            \includegraphics[width=\textwidth]{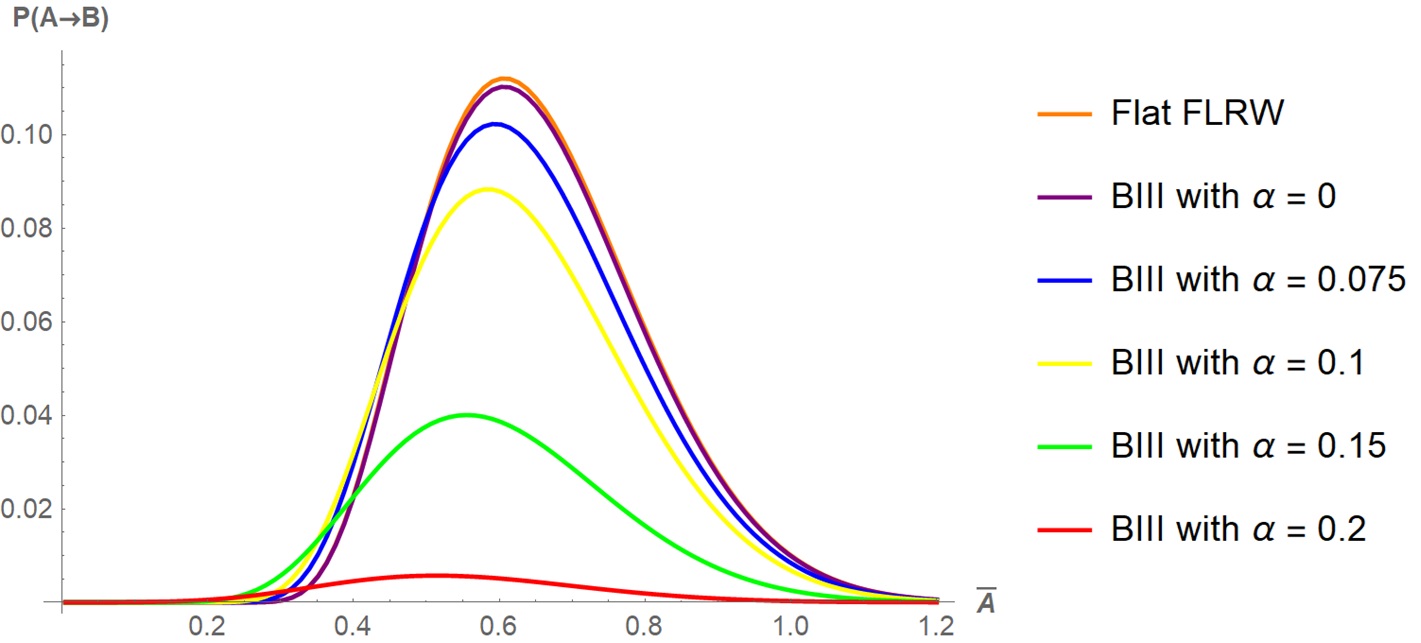}
                \caption{Transition probabilities for the flat FLRW (orange line) and Bianchi III metrics in units such that $\hbar=1$, choosing  $V_{B}=1$, $V_{A}=5$, $\frac{V^{(2)}_{B}}{V_{B}}-\frac{11}{3}=-0.005$, $T_{0}=1$, $T_1=0.1$, $T_2=0.05$, $(V_{A}^{(3)})^2=(V_{B}^{(3)})^2=0.1$, $V_{A}^{(4)}-3V_{A}^{(2)}=V_{B}^{(4)}-3V_{B}^{(2)}=0.1$. We plot the results of the semiclassical contribution plus the first and second quantum corrections, $\Gamma_0+\Gamma_1+\Gamma_2$, for different values of $\alpha$; we choose the cases $\alpha=0$ (purple line), $\alpha=0.075$ (blue line), $\alpha=0.1$ (yellow line), $\alpha=0.15$ (green line) and $\alpha=0.2$ (red line). The isotropy limit of the Bianchi III metric $\alpha \to 0$, does not correspond exactly to the flat FLRW result.}
            \label{fig:SQCBIII}
        \end{figure}

Now, by considering different values of $\alpha$, and arduously exploring the expression (\ref{eq:HIII}) we can find values of the parameters in which $H_{III}$ only takes real values. In this way, the first term in (\ref{eq:gamma2BIII}) does not contribute to the transition probability, we only have the tension term. This can be done for example by choosing $V_{B}=1$, $V_{A}=5$, $\frac{V^{(2)}_{B}}{V_{B}}-\frac{11}{3}=-0.005$, $T_{0}=1$, $T_1=0.1$, $T_2=0.05$, $(V_{A}^{(3)})^2=(V_{B}^{(3)})^2=0.1$, $V_{A}^{(4)}-3V_{A}^{(2)}=V_{B}^{(4)}-3V_{B}^{(2)}=0.1$.  The resulting transition probabilities are shown in Figure \ref{fig:SQCBIII} for different values of $\alpha$. It is shown that the general form corresponds to the behaviour encountered in the FLRW metrics. The effect of anisotropy is once again to reduce the probability and move the maximum value towards smaller values of $\bar{A}$ as anisotropy is increased.

In summary, we have obtained that the behaviour of the transition probabilities for the Bianchi III metric is, in general, the same as for the FLRW metrics, that is, the semiclassical contribution leads to a maximum value at the origin of $\bar{A}$, the first quantum correction reduces the probability but does not change the overall shape and the second quantum correction leads to a vanishing probability in this point. However, as we pointed out earlier for this metric this point does not describe a spatial singularity. What is being avoided in this case is the situation where one dimension is absent and the other two are infinite in size. We could try different factorizations looking for a scenario that allows us to have access to a spatial singularity in this anisotropic metric looking for a description where the second quantum correction avoids such singularity. However, as we will show in the Appendix \ref{S-AP} this scenario does not exist. Therefore, we conclude that the avoidance of the initial singularity due to quantum corrections is only possible in the isotropic case. Moreover, we have shown that the effect of the anisotropy is to reduce the probability in all scenarios, making it fall faster. In particular, when considering up to second quantum correction terms, the maximum point of the probability is moved towards smaller values of $\bar{A}$.
\section{Transitions for the Kantowski-Sachs metric}
\label{S-KS}
Finally, we will consider the Kantowski-Sachs metric. This metric describes an ani\-so\-tro\-pic universe or the interior of a Schwarzschild black hole. Following the parametrization proposed by Misner \cite{Misner} and employing a change of variables \cite{Garcia-Compean:2021syl} we can write this metric in the form
      \begin{equation}
       	d s^{2}=-N^{2}(t) d t^{2}+\gamma^{2}(t) d r^{2}+\frac{\sigma^{2}(t)}{\gamma^{2}(t)}\left[d \theta^{2}+\sin ^{2} \theta d \psi^{2}\right],
       	\label{MetricKS}
       \end{equation}
with $0 \leq \theta \leq \pi$ and $0 \leq \psi \leq 2 \pi$.  
Considering once again a homogeneous scalar field the Hamiltonian constraint turns out to be
        \begin{equation}
            H=N\left[\frac{\gamma^{3}}{4 \sigma^{2}} \pi_{\gamma}^{2}-\frac{\gamma}{4} \pi_{\sigma}^{2}+\frac{\gamma}{2 \sigma^{2}} \pi_{\phi}^{2}+\frac{\sigma^{2}}{\gamma} V(\phi)-\gamma\right] \simeq 0.
        \end{equation}
Once more, we can focus only on the term inside brackets, and comparing it with the general form (\ref{HamConst}), we observe that the coordinates defining the minisuperspace are $\left\{\Phi^{M}\right\}=\{\gamma, \sigma, \phi\}$. Thus, the metric  is
        \begin{equation}
            (G^{M N})=\left(\begin{array}{ccc}
            \frac{\gamma^3}{2 \sigma^2} & 0 & 0 \\
            0 & -\frac{\gamma}{2} & 0 \\
            0 & 0 & \frac{\gamma}{\sigma^2}
            \end{array}\right)
        \end{equation}
and the function $f$ is given by
        \begin{equation}
            f(\gamma, \sigma, \phi)=\frac{\sigma^{2}}{\gamma} V(\phi)-\gamma .
        \end{equation}
We also have in this case that the volume of the spatial slice $X$ reads
        \begin{equation}
            \operatorname{Vol}(X)=\int_{r} \int_{\theta=0}^{\pi} \int_{\psi=0}^{2 \pi} \sin \theta d r d \theta d \psi=4 \pi \int d r,
        \end{equation}
which is finite only if we restrict it to a finite interval in $r$, since the spatial slice is again non-compact in this scenario. Thus, we will consider it as an overall arbitrary constant as in the previous cases. 

In order to obtain a consistent system of equations we need to consider non-zero values for the potential minima. In the region of constant scalar field, the relations (\ref{Relations}) leads in this case to 
        \begin{equation}
            \gamma^{2}=\frac{\sigma^{2} V}{1-c \sigma},
        \end{equation}
where $c$ is an integration constant. By definition $\gamma$ and $\sigma$ are positive functions \cite{Garcia-Compean:2021syl}, therefore $c$ has to fulfill the condition
        \begin{equation}
            \frac{V}{1-c \sigma}>0, 
            \label{eq:ConstraintKS}
        \end{equation}
which depends on the value of $V_{A,B}$. From this expression we note that $\lim_{\sigma\to0}\gamma=0$ but $\lim_{\sigma\to0}\frac{\sigma^2}{\gamma^2}=\frac{1}{V}$. Therefore, we note from the form of the metric (\ref{MetricKS}) that the limit $\sigma\to0$ leads to the vanishing of the radial component but not of the angular part, thus, once again we will not have access to a spatial singularity.

Computing the remaining terms and substituting back in (\ref{eq:gammazeroVOL}) we obtain
	\begin{equation}
		\Gamma_0= \pm\frac{\operatorname{Vol}(X) i}{\hbar}\left[c_{B} F_{K S}\left[c_{B}, \sigma\right]\bigg\rvert_{\sigma_{0}}^{\bar{\sigma}}-c_{A}F_{K S}\left[c_{A}, \sigma\right]\bigg\rvert_{\sigma_{0}}^{\bar{\sigma}} \right]+\frac{\operatorname{Vol}(X)}{\hbar} \frac{\bar{\sigma} T}{\sqrt{V_{B}}} \sqrt{1-c_{B} \bar{\sigma}} ,
		\label{eq:G0KS}
	\end{equation}
where we have defined the function
        \begin{equation}
           \begin{aligned}
            F_{K S}[c, x]&=\int \frac{\sqrt{4-3 c x}}{1-c x} x d x \\
            &=-\frac{2}{9 c^{2}}\big[\sqrt{4-3 c x}(5+3 c x)-9 \operatorname{arctanh}(\sqrt{4-3 c x})\big].
            \end{aligned}
        \end{equation}
This result was first derived in \cite{Garcia-Compean:2021syl}. We note that $\lim_{\bar{\sigma}\to0}\Gamma_{0}=0$. Thus, once again the semiclassical contribution leads to a maximum value for the probability located at $\bar{\sigma}=0$. Furthermore, we note that in this semiclassical contribution, the tension term will be the only one that contributes after taking the real part since the integral with a constant scalar field gives always imaginary numbers. 

Moving further, for the first quantum correction we obtain
        \begin{equation}
            \nabla^2 f= V^{\prime\prime}.
        \end{equation}
Then, computing the remaining terms and substituting back in (\ref{eq:gamma1VOL}) we obtain for the first quantum correction
        \begin{equation}
            \Gamma_1= \frac{\operatorname{Vol}(X)}{2}\left[\left.\left(\frac{V_B^{(2)}}{V_B}-\frac{V_A^{(2)}}{V_A}\right) \ln (\sigma)\right|_{\sigma_0} ^{\bar{\sigma}-\delta \sigma}\right] + \operatorname{Vol}(X)T_{1},
        \end{equation}
where we note again that there only appears one tension term, thus, it only adds one extra parameter. Furthermore, we note that the divergence that appeared in the flat FLRW result in (\ref{eq:G1FLRFAux})  appears again here and we are led to impose also the condition (\ref{eq:ConditionsFirstQC}). Thus, we finally obtain 
	\begin{equation}
		\Gamma_1=\operatorname{Vol}(X)T_{1},
		\label{eq:G1KS}
	\end{equation}
which is only a constant. Therefore, the general behaviour is unaltered by the first quantum correction.

Carrying out with the analysis to the second quantum correction, we note that for this metric
        \begin{equation}
            \nabla^{2}\left(\nabla^{2} f\right)= \frac{\gamma  }{\sigma ^2}V^{(4)},
        \end{equation}
         \begin{equation}
           \left[\nabla\left(\nabla^{2} f\right)\right]^{2}= \frac{\gamma}{\sigma ^2}\left(V^{(3)}\right)^2 .
        \end{equation}
Therefore, computing the remaining necessary terms we obtain from (\ref{eq:gamma2VOL}) that the second quantum correction term is written as
	\begin{equation}
		\begin{aligned}
			\Gamma_2=\pm\frac{{\rm Vol}(X)i\hbar}{4} & \left\{ \frac{V_{B}^{(4)}}{V_{B}}H_{KS}(c_{B},\sigma)\bigg\rvert_{\sigma_{0}}^{\bar{\sigma}}+\left( \frac{V_{B}^{(3)}}{V_{B}}\right)^{2}I_{KS}(c_{B},\sigma)\bigg\rvert_{\sigma_{0}}^{\bar{\sigma}}-\frac{V_{A}^{(4)}}{V_{A}}H_{KS}(c_{A},\sigma)\bigg\rvert_{\sigma_{0}}^{\bar{\sigma}} \right. \\ & \left.-\left( \frac{V_{A}^{(3)}}{V_{A}}\right)^{2}I_{KS}(c_{A},\sigma)\bigg\rvert_{\sigma_{0}}^{\bar{\sigma}}\right\} +\frac{\operatorname{Vol}(X)\hbar}{\bar{\sigma}}\sqrt{\frac{V_{B}}{1-c_{B}\bar{\sigma}}}T_{2} ,
		\end{aligned}
		\label{eq:G2KS}
	\end{equation}
where we have defined the functions
        \begin{equation}
             H_{KS}(c,\sigma)=\int\frac{d\sigma}{\sigma^2\sqrt{4-3c\sigma}}=-\frac{\sqrt{4-3c\sigma}}{4\sigma}-\frac{3c}{8}\arctanh\left(\frac{\sqrt{4-3c\sigma}}{2}\right) ,
             \label{eq:MKS}
        \end{equation}
        \begin{equation}
            I_{KS}(c,\sigma)=\int\frac{1-c\sigma}{\sigma^2\left(4-3c\sigma\right)^{3/2}}d\sigma=-\frac{1}{16}\left[\frac{4-c\sigma}{\sigma\sqrt{4-3c\sigma}}+\frac{c}{2}\arctanh\left(\frac{\sqrt{4-3c\sigma}}{2}\right)\right] .
             \label{eq:NKS}
        \end{equation}

Furthermore, we note from this result that the functions $H_{KS}$ and $I_{KS}$ have a divergence if we choose  $\sigma_{0}=0$, but this divergence is independent of $c$. Therefore, in order to have a well defined probability we can eliminate those divergences by imposing the conditions
	\begin{equation}
		\frac{V^{(4)}_{B}}{V_{B}}=\frac{V^{(4)}_{A}}{V_{A}} , \hspace{1cm} \left(\frac{V_{B}^{(3)}}{V_{B}}\right)^2=\left(\frac{V_{A}^{(3)}}{V_{A}}\right)^2 .
	\end{equation} 

We note that as in all the previous cases, all the possible tension terms reduce to only one and that it is written in such a way that $\lim_{\bar{\sigma}\to0}\Gamma_{2}$ diverges, therefore choosing the appropriate signs the probability of having a universe created at $\bar{\sigma}=0$ vanishes. However, we note that as in the Bianchi III result, we are not avoiding the singularity, we are just preventing this point which has a different physical meaning. Furthermore, when we plot the transition probabilities for the semiclassical contribution in (\ref{eq:G0KS}), the first quantum correction in (\ref{eq:G1KS}) and the second quantum correction in (\ref{eq:G2KS}) we obtain the same qualitative behaviour as the one encountered for the FLRW metric in Figure \ref{fig:POSITIVEFLRWWC}. The study of the transition probabilities using this metric is relevant because although we can not compare directly the results with the Bianchi III or FLRW metric, in the sense that we can not obtain their results by taking an appropriate value of some parameter, we can see that the general behaviour that we encountered for the FLRW metric appears as well for this anisotropic metric but the point that gets avoided in this case does not describe a spatial singularity as the Bianchi III result. Thus, the conclusion that the singularity is avoided only in the isotropic universe is supported by the results of this section as well, showing that the limitation in the anisotropic case is not an artefact of the Bianchi III metric.

\section{Final Remarks}
\label{S-FinalRemarks}
\label{S-FinalR} 

In this article, we have studied the quantum corrections to the vacuum transition probabilities between two minima of a scalar field potential employing a Lorentzian formalism in the presence of gravity. First of all, we expanded a method previously proposed in \cite{Cespedes:2020xpn,Garcia-Compean:2021syl} to describe these probabilities by computing solutions of the WDW equation using a semiclassical expansion of the WKB type. We have shown how to take into account the quantum corrections to the general method for any model in the minisuperspace. We note that each quantum correction corresponds to considering one higher-order term in the $\hbar$ expansion of the WKB proposal. Then, the WDW equation provides one equation for every term in the semiclassical expansion considered. Therefore, in principle, the system of equations will always be solvable. Thus, this method can account for any desired order in the quantum corrections. Moreover, by choosing the parameter defining the integral curves in the minisuperspace in an appropriate form as has been used in previous works, and applying some general considerations, we have obtained explicit analytical expressions for the semiclassical contributions and up to second order quantum corrections in (\ref{eq:gammazeroVOL}), (\ref{eq:gamma1VOL}) and (\ref{eq:gamma2VOL}). One important feature of these general expressions is that all of them have the volume of the spatial slice as an overall constant, a feature that we expect will be maintained regardless of the order of the quantum correction that we wish to compute, and that allows to compute the transition probabilities for metrics with a non-compact spatial slice. Furthermore, all expressions are described in terms of one degree of freedom coming from the metric evaluated at the point where the transition is made. Thus, the interpretation of the probability functions obtained can still be stated as describing probability distributions of creating universes with a given size as was done for the semiclassical contributions on previous works. However, taking into account the quantum corrections, independent parameters (that we called tension terms) were added in increasing numbers. These terms arrived from the quantum fluctuations of the scalar field and may not be independent if the functions on the metric accompanying them are proportional or vanish. For example, on all the metrics considered in the main text, only one tension term was added for each order in the expansion. 

Then, we applied the method to some metrics of cosmological interest using General Relativity. We found analytical expressions for the transition amplitudes for homogeneous isotropic and anisotropic metrics. These expressions include the semiclassical contributions and two quantum corrections and are described by the potential minima and their derivatives, tension terms, and only one variable from the metric that represents the size of the universe at the time of nucleation.

For an isotropic universe, we studied the FLRW metric with positive and zero curvature. In order to obtain well defined probabilities, the quantum corrections implied extra restrictions on the derivatives of the potential evaluated at the minima. The general behaviour of the transition probabilities was shown in Figure \ref{fig:POSITIVEFLRWWC}. It was shown that the semiclassical contribution leads to a probability that starts at its maximum value on the spatial singularity and falls to zero as the scale factor increases. The first quantum correction does not change the overall shape, it only reduces the probability. On the other hand, the second quantum correction changes the behaviour in the singularity, leading to the prediction that the maximum probability is located at a small size of the scale factor but different from zero, in this way, the quantum corrections leads to an avoidance of the initial singularity. We also noted that the quantum correction terms are relevant in the ultraviolet described by small values of the scale factor but they do not change the behaviour in the infrared as we expected. It is interesting that the change of behaviour appears at second order in the quantum correction terms, the same occurs when computing quantum corrections to the Schrödinger equation coming from the Klein-Gordon equation as shown in \cite{Kiefer:1990pt}. Thus, since the WDW equation is a Klein-Gordon type equation in superspace we obtain a consistent result in this regard.

For an anisotropic universe, we studied the Bianchi III and the Kantowski-Sachs metrics. For the Bianchi metric, we encountered the same general form of the transition probability as in the isotropic case. However, the point of maximum probability in the semiclassical contribution that is avoided in the second quantum correction in this case does not represent an initial singularity. The different scale factors are related as a consequence of the integral curves on minisuperspace in such a way that such point describes a scenario where two Cartesian directions have infinite value for their scale factor whereas the third direction has a zero scale factor.  However, if we factorize some terms in the Hamiltonian constraint we can obtain new Hamiltonians that are equivalent at the classical level but at the quantum level they lead to different forms of the WDW equation and different relations of the scale factors. This ambiguity on the correct choice for the Hamiltonian constraint is equivalent to the ordering ambiguity in the quantization procedure for the squared momentum. However, as we showed in the Appendix \ref{S-AP}, once we obtain a proper factorization that allows an initial spatial singularity, the avoidance of such point is lost and we only obtain transition probabilities that lead to a maximum value at the initial singularity. Thus, the avoidance of the spatial singularity due to quantum corrections to the transition probabilities can only be obtained in the isotropic universe. Furthermore, at the classical level the limit $\alpha\to 0$ of the Bianchi III metric results in the flat FLRW metric. We encountered that the results from the semiclassical contribution indeed lead to the flat FLRW result in such isotropy limit. However, the quantum corrections lead to a very similar result but it differs from the flat FLRW result by some constants. Thus, the different structure of the minisuperspace from these two metrics is revealed only after considering quantum corrections. Furthermore, by varying the parameter $\alpha$ we found that the effect of anisotropy, even adding quantum corrections, is to decrease the transition probability, preserving the form that was observed for the FLRW case of positive curvature. On the other hand, for the Kantowski-Sachs metric we also encountered the same general behaviour for the transition probabilities as for the previous cases. But in this case, the point that gets avoided represents a point where the radial part of the metric is set to zero but not the angular part, thus, once again it is not a spatial singularity. Therefore, the result that the avoidance of the initial singularity is only possible for the isotropic universe is not an artefact of the Bianchi III metric.

Let us remark, that in the Euclidean formalism the computation of the quantum corrections to the transition probabilities is troublesome. However, in the Lorentzian formalism we could compute such corrections in a fairly simple form by considering higher order terms in the semiclassical expansion. It will be interesting to investigate if some of the properties outlined in this article such as the avoidance of the initial singularity for the isotropic universe can also be encountered on the Euclidean formalism or if this represents a relevant departure from both approaches. Moreover, it could also be relevant to pursue a possible generalization of the current method to midisuperspace models.

 \vspace{1cm}
\centerline{\bf Acknowledgements} \vspace{.5cm} J. H. Aguilar (No. 933342) would like to thank CONAHCyT for a grant. He also wants to thank Cinvestav for its hospitality. 

\appendix
\section{Effects of a given factorization to the transition probabilities}
\label{S-AP}
In order to determine the effect of choosing a different factorization for the Hamiltonian constraint in the Bianchi III and flat FLRW metrics, let us study a general factorization of the Hamiltonian of the Bianchi III metric (\ref{eq:HamConstB3}) in the form 
        \begin{equation}
            \begin{aligned}
           H=&N A^{-\gamma} B^{-\delta} C^{-\zeta}  \left[\frac{A^{\gamma+1} B^{\delta-1} C^{\zeta-1}}{4} \pi_{A}^{2}+\frac{B^{\delta+1} A^{\gamma-1} C^{\zeta-1}}{4} \pi_{B}^{2}+\frac{C^{\zeta+1} A^{\gamma-1} B^{\delta-1}}{4} \pi_{C}^{2}\right.\\ & -\frac{A^{\gamma} B^{\delta} C^{\zeta-1}}{2} \pi_{A} \pi_{B}-\frac{A^{\gamma} C^{\zeta} B^{\delta-1}}{2} \pi_{A} \pi_{C}-\frac{A^{\gamma-1} B^{\delta} C^{\zeta}}{2} \pi_{B} \pi_{C}+\frac{A^{\gamma-1} B^{\delta-1} C^{\zeta-1}}{2} \pi_{\phi}^{2} \\ & 
            \left.+\alpha^{2} A^{\gamma-1} B^{\delta+1} C^{\zeta+1}+V(\phi) A^{\gamma+1} B^{\delta+1} C^{\zeta+1}\right] \simeq 0,
            \end{aligned}
        \end{equation}
where $\gamma$, $\delta$ and $\zeta$ are integers. We now consider the Hamiltonian constraint to be the terms within brackets, then, after identifying the metric and the corresponding $f$ function we can proceed to relate the degrees of freedom of the metric when the scalar field is constant by employing eq. (\ref{Relations}). In this case we obtain
	\begin{equation}
		\frac{dA}{dB}=\left(\frac{A}{B}\right)\frac{(\gamma-1-\delta-\zeta)V+\frac{\alpha^2}{A^2}\left(\gamma-3-\delta-\zeta\right)}{(\delta-1-\gamma-\zeta)V+\frac{\alpha^2}{A^2}\left(\delta+1-\gamma-\zeta\right)} ,
	\end{equation}
and
        \begin{equation}
            \frac{d B}{d C}=\left(\frac{B}{C}\right) \frac{(\delta-1-\gamma-\zeta) V+\frac{\alpha^{2}}{A^{2}}(\delta+1-\gamma-\zeta)}{(\zeta-1-\gamma-\delta) V+\frac{\alpha^{2}}{A^{2}}(\zeta+1-\gamma-\delta)}.
        \end{equation}
The second relation is very difficult to solve because of the appearance of the $A$ function. We can simplify this expression by choosing $\delta=\zeta$. Then, we obtain
        \begin{equation}
            B=b_{0} C,
        \end{equation}
where $b_{0}$ is an integration constant that will be taken as 1. In the same way, it is found that 
        \begin{equation}
            B=a_{0} A^{\phi_{1}}\left[A^{2}+\phi_{2} \frac{\alpha^{2}}{V}\right]^{\phi_{3}},
            \label{eq:BdependA}
        \end{equation}
where $a_{0}$ is another integration constant that will be taken as 1 and
        \begin{equation}
            \phi_{1}=\frac{1-\gamma}{\gamma-3-2 \zeta}, \quad \phi_{2}=\frac{\gamma-3-2 \zeta}{\gamma-1-2 \zeta}, \quad \phi_{3}=\frac{2(\zeta+1)}{(2 \zeta+1-\gamma)(2 \zeta+3-\gamma)}.
        \end{equation}
Nevertheless, we want to obtain an isotropy limit that leads to the flat FLRW metric in order to correctly study the effect of anisotropy, that is we require $\lim_{\alpha=0} B=A$. This leads to the condition
        \begin{equation}
            \phi_{1}+2 \phi_{3}=1,
        \end{equation}
which gives us the solution $\zeta=\gamma$. In this way, the most general factorization requires $\gamma=\zeta=\delta$. So the above definitions are simplified as 
        \begin{equation}
            \phi_{1}=\frac{\gamma-1}{\gamma+3}, \quad \phi_{2}=\frac{\gamma+3}{\gamma+1}, \quad \phi_{3}=\frac{2}{\gamma+3} . 
        \end{equation} 
However, as we remarked in the main text the motivation to consider a different factorization is to have access to an initial singularity for the Bianchi III metric, that is that we can have $\lim_{A\to0}B=0$. We note from eq. (\ref{eq:BdependA}) that this behaviour can be obtained by requiring that $\phi_{1}>0$, which leads to
	\begin{equation}
		\gamma>1 , \hspace{0.5cm} \text{or} \hspace{0.5cm} \gamma<-3 .
		\label{eq:APConditionGamma}
	\end{equation}
	
Furthermore, the general form of the metric in minisuperspace is
        \begin{equation}
            \left(G^{M N}\right)=\frac{1}{2}\left(\begin{array}{cccc}
            A^{\gamma+1}(B C)^{\gamma-1} & -(A B)^{\gamma} C^{\gamma-1} & -(A C)^{\gamma} B^{\gamma-1} & 0 \\
            -(A B)^{\gamma} C^{\gamma-1} & B^{\gamma+1}(A C)^{\gamma-1} & -(B C)^{\gamma} A^{\gamma-1} & 0 \\
            -(A C)^{\gamma} B^{\gamma-1} & -(B C)^{\gamma} A^{\gamma-1} & C^{\gamma+1}(A B)^{\gamma-1} & 0 \\
            0 & 0 & 0 & 2(A B C)^{\gamma-1}
            \end{array}\right) 
        \end{equation}
and we can identify
        \begin{equation}
            f=H(A, B, C)+F(A, B, C) V(\phi),
        \end{equation}
with
        \begin{equation}
            H(A, B, C)=\alpha^{2} A^{\gamma-1}(B C)^{\gamma+1}, \quad F(A, B, C)=(A B C)^{\gamma+1}.
        \end{equation}
With these considerations, we can proceed to compute the transition probabilities with up to second quantum corrections using the general expressions (\ref{eq:gammazeroVOL}), (\ref{eq:gamma1VOL}) and (\ref{eq:gamma2VOL}). However, we note that all the integrals on the three expressions will vanish in the limit $\bar{A}\to A_{0}=0$. Therefore, the behaviour of the transition probabilities in the singularity is dictated only by the tension terms. Thus, let us analyse the resulting tension terms due to the three contributions.

Following (\ref{eq:gammazeroVOL}) the tension term for the semiclassical contribution takes the form 
        \begin{equation}
            \frac{\operatorname{Vol}(X)}{\hbar} \bar{A}^{\frac{(\gamma+1)(3 \gamma+1)}{\gamma+3}}\left[\bar{A}^{2}+\frac{\gamma+3}{\gamma+1} \frac{\alpha^{2}}{V_{B}}\right]^{\frac{4(\gamma+1)}{\gamma+3}} T_{0} .
            \label{eq:T0BIIIgamma}
        \end{equation}
From this expression, we note that the important quantity to keep track, to study the behaviour in the initial singularity, is given by
	\begin{equation}
		\beta_{0}=\frac{(\gamma+1)(3\gamma+1)}{\gamma+3} .
	\end{equation}
Furthermore, taking the limit $\alpha\to0$ of (\ref{eq:T0BIIIgamma}) we are lead consistently to the tension term of the flat FLRW metric
	\begin{equation}
		\frac{\operatorname{Vol}(X)}{\hbar} \bar{a}^{3(\gamma+1)} T_{0} ,
	\end{equation}
from where we note that relevant quantity is
	\begin{equation}
		\eta_{0}=3(\gamma+1) .
	\end{equation}
In the same way, following (\ref{eq:gamma1VOL}) for the first quantum correction $\Gamma_1$, we obtain only one tension term written as
        \begin{equation}
            \operatorname{Vol}(X) \bar{A}^{\frac{2\gamma(3 \gamma+1)}{\gamma+3}}\left[\bar{A}^{2}+\frac{\gamma+3}{\gamma+1} \frac{\alpha^{2}}{V_{B}}\right]^{\frac{8\gamma}{\gamma+3}} T_{1}.
            \label{eq:T1BIIIgamma}
        \end{equation}
In the previous equation (\ref{eq:T1BIIIgamma}) we see that the relevant quantity to study the initial singularity is
		\begin{equation}
			\beta_{1}=\frac{2\gamma(3 \gamma+1)}{\gamma+3} .
		\end{equation}
One again, taking the limit $\alpha\to0$, we obtain the corresponding tension term for the flat FLRW metric
	\begin{equation}		
		{\rm Vol}(X) \bar{a}^{6 \gamma }T_{1} ,     
	\end{equation}
where the relevant quantity is
	\begin{equation}
		\eta_{1}=6\gamma .
	\end{equation}
  
Finally, for the second quantum correction (\ref{eq:gamma2VOL}), we obtain only 3 different tension terms in the form
        \begin{equation}
             \begin{aligned}
            \operatorname{Vol}(X)\hbar\left\{\bar{A}^{\frac{9 \gamma^{2}-1}{\gamma+3}}\left[\bar{A}^{2}+\frac{\gamma+3}{\gamma+1} \frac{\alpha^{2}}{V_{B}}\right]^{\frac{4(3 \gamma-1)}{\gamma+3}} T_{2,1}+\bar{A}^{\frac{15 \gamma^{2}-7}{\gamma+3}}\left[\bar{A}^{2}+\frac{\gamma+3}{\gamma+1} \frac{\alpha^{2}}{V_{B}}\right]^{\frac{4(5 \gamma-1)}{\gamma+3}} T_{2,2}\right. \\
           + \left.\bar{A}^{\frac{(5 \gamma-1)(3 \gamma+1)}{\gamma+3}}\left[\bar{A}^{2}+\frac{\gamma+3}{\gamma+1} \frac{\alpha^{2}}{V_{B}}\right]^{\frac{4(5 \gamma-1)}{\gamma+3}} T_{2,3}\right\}.
            \end{aligned}
            \label{eq:T2BIIIgamma}
        \end{equation}
Therefore, the relevant quantities for the second correction are
	\begin{equation}
		\beta_{2}=\frac{9 \gamma^{2}-1}{\gamma+3} , \hspace{0.5cm} \beta_{3}=\frac{15 \gamma^{2}-7}{\gamma+3} , \hspace{0.5cm} \beta_{4}=\frac{(5 \gamma-1)(3 \gamma+1)}{\gamma+3} .
	\end{equation}
In this case, in order to obtain correctly the FLRW flat result, we need to take  $\alpha\to0$ and eliminate the second tension term since the functions depending on the scalar field vanish in this limit. Thus, we are led to 
	\begin{equation}
		 \operatorname{Vol}(X)a^{3(3\gamma-1)}T_{2,1}+\operatorname{Vol}(X)a^{3(5\gamma-1)}T_{2,3} ,
	\end{equation} 
from where we identify
	\begin{equation}
		\eta_{2}=3(3\gamma-1)  , \hspace{0.5cm} \eta_{3}=3(5\gamma-1) .
	\end{equation}

Therefore, the singularity will be the point of maximum probability for $\beta_{i}$ and $\eta_{i}$ positive, the general form of the probability distribution will not change for zero values of these parameters, whereas the singularity will be avoided if at least one of the $\beta_{i}$ (for the Bianchi III metric) or one of the $\eta_{i}$ (for the flat FLRW metric) is negative. 

As asserted before we are interested in studying these behaviours for the regions where we can explore the initial singularity, that is $\gamma>1$ or $\gamma<-3$. However, it can be shown that all $\beta_{i}$ and $\eta_{i}$ are positive for $\gamma>1$ whereas all of these coefficients are negative for $\gamma<-3$. Since taking negative values for $\eta_{0}$ will change drastically our semiclassical results, we will not consider the region $\gamma<-3$. Therefore, focusing on $\gamma>1$ we note that all coefficients are positive, thus, the singularity will remain as the point with maximum probability even after taking up to second quantum correction terms. Therefore, we conclude that the avoidance of the initial singularity by taking quantum corrections can only be performed in the isotropic universe.


\end{document}